\documentclass[12pt]{article}
\usepackage{amssymb,amscd,amsmath,amsthm}
\usepackage{latexsym,amstext}
\usepackage{latexsym,amstext}
\usepackage{color}
\usepackage{dcolumn}
\usepackage{bm}
\usepackage{cite}

\usepackage{graphicx}
\usepackage{epsfig,epstopdf}

\def\ds{\displaystyle}
\def\be{\begin{equation}}
\def\ee{\end{equation}}

\def\bea{\begin{eqnarray}}
\def\eea{\end{eqnarray}}
\def\bean{\begin{eqnarray*}}
\def\eean{\end{eqnarray*}}

\begin{document}

\title{\bf Groups, Special Functions and Rigged Hilbert Spaces}

\author{E. Celeghini$^{1,2}$\footnote{E.mails: celeghini@fi.infn.it\,;
manuelgadella1@gmail.com\,;
marianoantonio.olmo@uva.es \,.
}, M. Gadella$^2$, M. A. del Olmo$^2$\\ \\
$^1$ Dipartimento di Fisica, Universit\`a di Firenze and\\  INFN-Sezione di Firenze\\
150019
Sesto Fiorentino, Firenze, Italy\\
$^2$ Departamento de F\'{\i}sica Te\'orica, At\'omica y Optica  and IMUVA, \\
Universidad de Va\-lladolid, 47011 Valladolid, Spain\\
}

\maketitle

\begin{abstract}{
We show that Lie groups and their respective algebras, special functions and rigged Hilbert spaces are complementary concepts that coexist together in a common framework and that they are aspects of the same mathematical reality. Special functions serve as bases for infinite dimensional Hilbert spaces supporting linear unitary irreducible  representations of a given Lie group. These representations are explicitly given by operators on the Hilbert space $\mathcal H$ and the generators of the Lie algebra are represented by unbounded self-adjoint operators. The action of these operators on elements of continuous bases is often considered. These continuous bases do not make sense as vectors in the Hilbert space, instead they are functionals on the dual space, $\Phi^\times$, of a rigged Hilbert space, $\Phi\subset \mathcal H \subset \Phi^\times$. As a matter of fact, rigged Hilbert spaces are the structures in which both, discrete orthonormal and continuous bases may coexist. We define the space of test vectors $\Phi$ and a topology on it at our convenience, depending on the studied group. The generators of the Lie algebra can be often continuous operators on $\Phi$ with its own topology, so that they admit continuous extensions to the dual $\Phi^\times$ and, therefore, act on the elements of the continuous basis.  We have investigated this formalism to various examples of interest in quantum mechanics.  In particular, we have considered, $SO(2)$ and functions on the unit circle, $SU(2)$ and associated Laguerre functions, Weyl-Heisenberg group and Hermite functions, $SO(3,2)$ and spherical harmonics, $su(1,1)$ and Laguerre functions, $su(2,2)$ and algebraic Jacobi functions and, finally, $su(1,1)\oplus su(1,1)$ and Zernike functions on a circle. }
\end{abstract}\bigskip



\section{Introduction}\label{introduction}

Harmonic analysis has undergone strong development  since  the first  work by Fourier  \cite{fourier}. The main idea of the Fourier method is to decompose functions in a superposition of other particular functions, i.e, ``special functions''.  Since the original trigonometric functions  used by Fourier many special functions, like the classical orthogonal polynomials  \cite{folland92},  have been used generalising the original Fourier idea. In many cases such special functions support representations of groups and in this way group representation theory appears closely linked to  harmonic analysis \cite{folland95}.  Another cruzial fact is that harmonic analysis  is related to  linear algebra and functional analysis, in the sense that  elements of vector spaces or Hilbert spaces are decomposed in terms of  orthogonal bases  or operators as linear combinations of their eigenvalues (i.e  applying the spectral theorem, see Theorem 1 in Section \ref{riggedhs}).  In many occasions continuous bases and discrete bases are involved in the same framework, Hence the arena where all these objects fit in a   precise mathematical way is inside a rigged Hilbert space.
Hence we have a set of mathematical objects: classical orthogonal polynomials, Lie algebras, Fourier analysis, continuous and discrete bases and rigged Hilbert spaces fully incorporated in a  harmonic frame that can bee used in quantum mechanics as well as in signal processing. 

In a series of previous articles, we gave some examples showing that Lie groups and algebras, special functions, discrete and continuous bases and rigged Hilbert spaces (RHS) are particular aspects of the same mathematical reality, for which a general theory is needed. As a first step in the construction of this general theory, we want to present a compact review of the results which have been so far obtained by us and that can be useful in applications where harmonic analysis is involved. 

Special functions play often the role of being part of orthonormal bases of Hilbert spaces serving as support of representations of Lie groups of interest in Physics. As is well known, decompositions of vectors of these spaces are given in terms of some sort of {\it continuous basis}, which are not normalisable and hence, outside the Hilbert space. The most popular formulation to allow the coexistence of these continuous bases with the usual discrete bases is the RHS, where the elements of continuous bases are well defined as functionals on a locally convex space densely defined as a subspace of the Hilbert space supporting the representation of the Lie group.

Thus, we have the need for a framework  that includes  Lie algebras, discrete and continuous bases and special functions, as building blocks of these discrete bases. In addition, it would be desirable to have structures in which the generators of the Lie algebras be well defined continuous operators on.  The  
rigged Hilbert space comply with the requirements above mentioned. 

All the cases presented here have applications not only in physics but in other sciences. In particular,  Hermite functions are related to signal analysis in the real line and also with the fractional Fourier transform \cite{OZA}. In \cite{CGO19} we have introduced a new set of functions in terms of the Hermite functions that give rise bases in $L^2(\mathcal C)$ and in 
$l^2(\mathbb Z)$ where $\mathcal C$ is the unit circle. Both bases are related by means of the Fourier transform and the discrete Fourier transform. In \cite{CGO20} we will present a systematic study of these functions as well as the corresponding rigged Hilbert space framework.   Recently spherical harmonics are used in 3-dimensional signal processing with applications in geodesy, astronomy, cosmology, graph computation, vision computation, medical images, communications systems,... \cite{KS,RH,MRC}. Zernike polynomials are well known for their applications in optics \cite{ZER,mahajana2010,lakshminarayanana2011}. 
Moreover all of them can be considered as examples of harmonic analysis where the connection between groups, special functions and RHS fit together perfectly.

The paper is  organized as follows. A brief description of RHS and their use in Physics and in Engineering is given in Section \ref{riggedhs}. 
In Section \ref{so2preliminary} it is discussed in details SO(2), related to the exponential $e^{{\bf i}m\phi}$, where the technical aspects are reduced to the minimum. Section \ref{su2associatedlaguerre}  considers how Associated Laguerre Functions allow to construct two different RHS, one related to the integer spin and the other to half-integer spin of SU(2). In Section \ref{weylheisengerghermite} an analysis is performed of the basic case of the line, where the 
fundamental ingredient of the RHS are the Fourier Transform, the Hermite Functions and the Weyl-Heisenberg group. Section \ref{so32sphericalharmonics} is devoted to the RHS constructed on Spherical Harmonics in relation with SO(3). In section \ref{su11laguerre} Laguerre Functions are used to construct another RHS related to SU(1,1).
Jacobi Functions and the 15-dimensional algebra SU(2,2) are the bricks of a more large RHS which is studied in Section \ref{algebraicjacobifunctions}. The last example we discuss (Section \ref{zernikeRHS}) is the RHS constructed on the Zernike Functions and the algebra $su(1,1) \oplus su(1,1)$ (that should be used also in connection with the Laguerre Functions). Few remarks close the paper in Section \ref{conclusions}.


\section{Rigged Hilbert Spaces}\label{riggedhs}

The less popular among our ingredients is the concept of rigged Hilbert spaces, so that a short section devoted to this concept seems necessary. A rigged Hilbert space, also called Gelfand triplet, is a tern of spaces \cite{GEL}
\begin{equation}\label{1}
\Phi\subset\mathcal H\subset\Phi^\times\,,
\end{equation}
where: i.) $\mathcal H$ is an infinite dimensional separable Hilbert space; ii.) $\Phi$ is a dense subspace of $\mathcal H$ endowed with a locally convex topology stronger, i.e., it has more open sets, than the Hilbert space topology that $\Phi$ has inhereted from $\mathcal H$; iii.) $\Phi^\times$ is the space of all continuous antilinear functionals on $\Phi$. Thus, $F\in\Phi^\times$ is a mapping $F:\Phi\longmapsto \mathbb C$ such that  for any pair $\psi,\varphi\in\Phi$ and any pair of complex numbers $\alpha,\beta\in\mathbb C$, one has 
\begin{equation}\label{2}
F(\alpha\psi+\beta\varphi)=\alpha^*F(\psi)+\beta^*F(\varphi)\,,
\end{equation}
where the star denotes complex conjugation. The continuity is given with respect to the locally convex topology on $\Phi$ and the usual topology on the complex plane $\mathbb C$. Instead the notation in \eqref{2}, we shall henceforth use the Dirac notation, which is quite familiar to physicists:
\begin{equation}\label{3}
F(\varphi)=:\langle\varphi|F\rangle\,.
\end{equation}

In general, the topology on $\Phi$ is given by a family of seminorms. In the examples we have studied so far, the topology on $\Phi$ is given by a countable set of seminorms, where by countable we mean either finite or denumerable. 
As the topology on $\Phi$ is stronger than the Hilbert space topology, one of these seminorms could be chosen to be the Hilbert space norm. 

Seminorms provide a nice criterion to determine whether a linear or antilinear functional over $\Phi$ is continuous. The linear or antilinear functional $F:\Phi\longmapsto\mathbb C$ is continuous if and only if, there exists a positive number $K>0$ and finite number of seminorms, $p_1,p_2,\dots,p_m$, taken from those that define the topology on $\Phi$ such that for any $\varphi\in\Phi$, we have \cite{RSI}
\begin{equation}\label{4}
|\langle\varphi|F\rangle|\le K \{p_1(\varphi)+p_2(\varphi)+\dots+p_m(\varphi)\}\,.
\end{equation}
One typical example of functional is the following one. Pick an arbitrary $\varphi\in\Phi$ and define $F_\varphi$ as \
\begin{equation}\label{4bis}
F_\varphi(\psi)=\langle\psi|F_\varphi\rangle:= \langle\psi|\varphi\rangle\,,
\end{equation}
 which is obviously antilinear on $\Phi$. Then, use the Schwarz inequality in $|\langle\psi|F_\varphi\rangle|\le ||\psi||\,||\varphi||  $, take $K=||\varphi||$, $p_1(\psi):=||\psi||$, for all $\psi\in\Phi$ and use \eqref{4} so as to conclude the continuity of $F_\varphi$ on $\Phi$. However, not all elements of $\Phi^\times$ lie in this category. A typical counterexample is the Dirac delta.  

Analogously, assume that $\Phi$ and $\Psi$ are two locally convex spaces with topologies given by the respective families of seminorms $\{p_{i}\}_{i\in \mathcal I}$ and $\{q_{j}\}_{j\in \mathcal J}$. A linear or antilinear mapping $F:\Phi\longmapsto\Psi$ is continuous if and only if for each seminorm $q_j$ on $\Psi$ there exists a positive constant $K>0$ and a finite number of seminorms, $p_1,p_2,\dots,p_m$, from those defining the topology on $\Phi$ such that
\begin{equation}\label{5}
q_j(F(\varphi)) \le K \{ p_1(\varphi)+p_2(\varphi)+\dots+p_m(\varphi)\}\,,\qquad \forall \varphi\in \Phi\,.
\end{equation}
 Both, the constant $K$ and the seminorms $p_1,p_2,\dots,p_m$ depend on $q_j$, but not on $\varphi$. We shall use these results along the present article. 

Less interesting is that the dual space $\Phi^\times$ may be endowed with the weak topology induced by $\Phi$. As is well known, the seminorms for this weak topology are defined as follows: for each $\varphi\in\Phi$, we define the seminorm $p_\varphi$  as $p_\varphi(F)=|\langle\varphi|F\rangle|$, for all $F\in\Phi^\times$. 

Since the topology on $\Phi$ is stronger than the Hilbert space topology, the canonical injection $i:\Phi\longmapsto\mathcal H$, with $i(\varphi)=\varphi$, for all $\varphi\in\Phi$, is continuous. Furthermore, one may prove that the injection $i:\mathcal H\longmapsto \Phi^\times$ given by $i(\varphi):=F_\varphi$ \eqref{4bis} is one-to-one and continuous with respect to the Hilbert space topology on $\mathcal H$ and the weak topology on $\Phi^\times$ \cite{GEL}. 

RHS have been introduced in Physics with the purpose of giving a rigorous mathematical background to the celebrated Dirac formulation of quantum mechanics, which is widely used by  physicists.  This mathematical formulation has been the objects of various publications \cite{BOHM2,RO,ANT,MEL,GG,GG1,GG2}.  In addition, rigged Hilbert spaces have been used in Physics or mathematics with various purposes that include:

1. A proper mathematical meaning for the Gamow vectors, which are the non-normalizable vectors giving the states of the exponentially decaying part of a quantum scattering resonance \cite{BOHM,BOHM1,BG,CG}.

2. Using Hardy functions on a half-plane \cite{BG,CG,GAD}, we may construct RHS that serve as a framework for an extension of ordinary quantum mechanics that accounts for time asymmetric quantum processes. One example of such processes is the quantum decay \cite{ARN,BHKW,BLV,BKW,BKW1,BGK}. 

3. Providing an appropriate context for the spectral decompositions of Koopman and Frobenius-Perron operators in classical chaotic systems in terms of the so called Pollicot-Ruelle resonances, which are singularities of the power spectrum \cite{AT,AT1}.

4. Some situations that arise in quantum statistical mechanics demand  the use of generalized states and some singular structures that require the use of rigged Liouville spaces \cite{AGS,AGS1,AG}.

5. A proper definition of some of the structures that appear in the axiomatic theory of quantum fields, like Wightman functional, Borchers algebra, generalized states, etc, require of structures like the rigged Fock space \cite{BLT,AGPP,GAD6}. Both rigged Liouville and Fock spaces are obvious generalizations of RHS. 

6. White noise and other stochastic processes may also be formulated in the context of RHS \cite{HID,HID1} as well as the study of certain solutions of partial differential equations \cite{HOR}. 

7. In the last years the RHS have appeared associated to time-frequency analysis and Gabor analysis that have many applications in physics and engineering related to signal proceesing 
\cite {fezim98,cofelu08,fe09,ban10,feja18,fe18-3,feja19,feluja19}.  In particular, applications in electrical engineering   have been introduced in \cite{HJ1,HJ2,HJ3}. 

One of the most interesting properties of RHS is the possibility of extending to the duals certain unbounded operators defined on domains including the space $\Phi$. Let us consider a linear operator  $A: \Phi\longmapsto\mathcal H$ and let $A^\dagger$ its adjoint, which has the following properties: 

1. For any $\varphi\in\Phi$, then, $A^\dagger\varphi\in\Phi$. One says that $\Phi$ reduces $A^\dagger$ or, equivalently, that $A^\dagger$ leaves $\Phi$ invariant, so that $A^\dagger\Phi\subset\Phi$. Note that we do not assume that $A\Phi\subset\Phi$, in general. 

2. The adjoint $A^\dagger$ is continuous on $\Phi$. 

Then, the operator $A$ may be extended to a continuous operator on $\Phi^\times$, endowed with the weak topology. For any $F\in\Phi^\times$, one defines $AF$ as by means of the following {\it duality formula}:   
\begin{equation}\label{6}
\langle \varphi|AF\rangle=\langle A^\dagger\varphi|F\rangle\,,\qquad \forall\,\varphi\in\Phi\,,\;\; \forall\,F\in\Phi^\times\,.
\end{equation}
One of the properties of the extension is that it is continuous on $\Phi^\times$ when this space has the weak topology. 

In particular, if $A$ is a symmetric operator, formula \eqref{6} read as $\langle \varphi|AF\rangle=\langle A\varphi|F\rangle$. If $A$ were self-adjoint, there is always a subspace $\Phi$ with the following properties: i.) $\Phi$  dense in $\mathcal H$; ii.) $\Phi$ is a subspace of the domain of $A$; iii.) it is possible to endow $\Phi$ with a locally convex topology, finer than the Hilbert space topology, such that $A$ be continuous on $\Phi$. As a consequence, there exists a RHS, $\Phi\subset\mathcal H\subset\Phi^\times$ such that the self-adjoint operator $A$ may be extended to the dual $\Phi^\times$ and, henceforth, to a larger space than the original Hilbert space where $A$ is densely defined. 

These ideas drive us to the important result known as the Gelfand-Maurin theorem \cite{GEL,MAU} that gives a spectral decomposition \`a la Dirac of a self-adjoint operator with continuous spectrum. We present it here in its simplest form in order not to enter in unnecessary complications and notations.\medskip

{\bf Theorem 1.-}(Gelfand-Maurin) {\it
Let $A$ be a self-adjoint operator on a infinite dimensional separable Hilbert space $\mathcal H$, with simple absolutely continuous spectrum $\sigma(A)\equiv \mathbb R^+\equiv [0,\infty)$. Then, there exists a rigged Hilbert space $\Phi\subset \mathcal H\subset\Phi^\times$, such that:

1. $A\Phi\subset\Phi$ and $A$ is continuous on $\Phi$. Therefore, it may be continuously extended to $\Phi^\times$.

2. For almost all $\omega\in \mathbb R^+$, with respect to the Lebesgue measure, there exists a $|\omega\rangle\in\Phi^\times$ with $A|\omega\rangle = \omega\,|\omega\rangle$.

3. (Spectral decomposition)  For any pair of vectors $\varphi,\psi\in\Phi$, and any measurable function $f:\mathbb R^+\longmapsto \mathbb C$, we have that
\begin{equation}\label{7}
\langle\varphi|f(A)\psi\rangle= \int_0^\infty f(\omega)\, \langle \varphi|\omega\rangle\langle \omega|\psi\rangle\,d\omega\,,
\end{equation}
with $\langle\omega|\psi\rangle=\langle\psi|\omega\rangle^*$. 

4. The above spectral decomposition is implemented by a unitary operator $U:\mathcal H \longmapsto L^2(\mathbb R^+)$, with $U\psi=\langle \omega|\psi\rangle=\psi(\omega)$ and $[UAU^{-1}]\psi(\omega)=\omega \,\psi(\omega)=\omega\langle\omega|\psi\rangle$ for any $\psi\in\Phi$ . This means that $UAU^{-1}$ is the multiplication operator on $U\Phi$. 

5. For any pre-existent RHS $\Phi\subset \mathcal H \subset \Phi^\times$, such that $A\Phi\subset\Phi$ with continuity and $A$ is an operator satisfying our hypothesis, then 2, 3 and 4 hold. 
}

This result will have some interest in our forthcoming discussion. 

Two rigged Hilbert spaces $\Phi\subset\mathcal H\subset \Phi^\times$ and $\Psi\subset\mathcal G\subset\Psi^\times$ are unitarily equivalent if there exists a unitary operator $U:\mathcal H\longmapsto\mathcal G$ such that: i.) $U$ is a {\it one-to-one}  mapping from $\Phi$ {\it onto} $\Psi$; ii.) $U:\Phi\longmapsto \Psi$ is continuous; iii.) its inverse $U^{-1}:\Psi\longmapsto \Phi$ is also continuous. Then, by using the duality formula
\begin{equation}\label{8}
\langle \varphi|F\rangle= \langle U\varphi|UF\rangle\,, \qquad \forall\,\varphi\in\Phi\,,\;\; \forall \,F\in\Phi^\times\,,
\end{equation}
we extend $U$  to a one-to-one mapping from $\Phi^\times$ onto $\Psi^\times$, which is continuous with the weak topologies on both duals and which has an inverse with the same properties. Resumming we have the following diagram
\[
\begin{array}{lllclll}
&\Phi&\subset &{\mathcal H}&\subset&\Phi^\times
\\[0.2cm]
  & \hskip-0.35cm  {U}\downarrow && \hskip-0.30cm {U}\downarrow &&\hskip-0.30cm {U}\downarrow
\\[0.2cm]
&\Psi &\subset  & \mathcal G &\subset   &\Psi^\times
\end{array}\,.
\]


\section{SO(2): The basic example}\label{so2preliminary}

To begin with, let us briefly summarize the most simple case that contains some general ingredients to be used in other situations \cite{CGO}.  Consider the unit circle in the plane, defined  by 
 $\mathcal C:= \{(x,y)\in\mathbb R^2\;;\; x^2+y^2=1\}$. As is well known, its group of invariance is $SO(2)$. 
 
 The Hilbert space on the unit circle is the space of measurable functions $f(\phi):\mathcal C\longmapsto \mathbb R$, which are square integrable. We denote this space as $L^2[0,2\pi)\equiv L^2(\mathcal C)$. The set of functions  
 \begin{equation}\label{9aa} 
 f_m(\phi):=\frac1{\sqrt{2\pi}}\,e^{-im\phi}\,,\qquad \forall m\in\mathbb Z\,,  
 \end{equation}
   where $\mathbb Z$ is the set of integer numbers, is an orthonormal basis in $L^2[0,2\pi)$. Then, each $f(\phi)\in L^2(\mathcal C)$ admits a span of the form,
\begin{equation}\label{9}
f(\phi)= \sum_{m\in\mathbb Z} a_m\,f_m(\phi)\,,\qquad a_m\in \mathbb C
\end{equation}
 with
 \begin{equation}\label{9a}  
 \sum_{n\in\mathbb Z} |a_m|^2=\big|\big|f(\phi)\big|\big|^2\,,
\end{equation}
where $\mathbb C$ is the  field of the complex numbers and $\big|\big|f(\phi)\big|\big|$ is the norm of the function $f(\phi)$ on $L^2[0,2\pi)$. 
\subsection{Rigged Hilbert spaces associated to $L^2(\mathcal C)$}

To construct a RHS, let us consider the space $\Psi$ of the functions  $f(\phi)\in L^2[0,2\pi)$ having the property,
\begin{equation}\label{10}
\big|\big|f(\phi)\big|\big|_p^2:= \sum_{m\in\mathbb Z} |a_m|^2\,|m+i|^{2p}<\infty\,,\qquad p=0,1,2\dots\,.
\end{equation}
The countably family of norms $\big|\big|-\big|\big|_p$ generates a metrizable topology on $\Psi$. The fact that this family includes $p=0$, shows that the canonical injection $\Psi\longmapsto L^2(\mathcal C)$ is continuous. Let $\Psi^\times$ be the dual of $\Psi$ (continuous antilinear functionals on $\Psi$) with the weak topology induced by the dual pair $\{\Psi,\Psi^\times\}$. Then, $\Psi\subset L^2(\mathcal C)\subset \Psi^\times$ is a RHS. 

Along this particular and concrete RHS, we consider another one, unitarily equivalent to this and constructed as follows. Let us take an abstract infinite dimensional separable Hilbert space $\mathcal H$. We know that there is a unitary  mapping $U\mathcal H\equiv L^2(\mathcal C)$, in fact continuous. The sequence of vectors $\{|m\rangle\}_{m\in\mathbb Z}$, with $U|m\rangle = f_m(\phi)$, forms a orthonormal basis on $\mathcal H$. Then, following the comment at the end of Section 2, we may construct a RHS, $\Phi\subset\mathcal H\subset\Phi^\times$ unitarily equivalent to $\Psi\subset L^2(\mathcal C)\subset \Psi^\times$, just by defining $\Phi:=U^{-1}\Psi$ and extending $U^{-1}$ as a continuous mapping from $\Psi^\times$ onto $\Phi^\times$, using the duality formula \eqref{8}. More explicitly
\[
\begin{array}{lllclll}
&\Psi&\subset &L^2(\mathcal C)&\subset&\Psi^\times
\\[0.2cm]
  & \hskip-0.74cm  {U^{-1}}\downarrow && \hskip-0.675cm {U^{-1}}\downarrow 
  &&\hskip-0.71cm {U^{-1}}\downarrow
\\[0.2cm]
&\Phi &\subset  & {\mathcal H} &\subset   &\Phi^\times
\end{array}\,.
\]
The mapping $U^{-1}$ also transport  topologies, so that if $|f\rangle\in\Phi$ with 
\begin{equation}\label{11}
|f\rangle=\sum_{m\in\mathbb Z} a_m\,|m\rangle =\sum_{m=-\infty}^\infty a_m\,|m\rangle\,,
\end{equation}
 then the topology on $\Phi$ is given by the set of norms $\big|\big|\,|f\rangle\big|\big|_p^2 =\sum_{m\in\mathbb Z} |a_m|^2\,(m+i)^{2p}$. 

One of the most important features of RHS is the possibility  of using continuous and discrete bases within the same space. For  any $\phi\in[0,2\pi)$, we define the ket $|\phi\rangle$ as a linear mapping from $\Phi$ into  $\mathbb C$, such that for any $|f\rangle\in\Phi$, with $|f\rangle=\sum_{m\in\mathbb Z} a_m\,|m\rangle$, we have 
\begin{equation}\label{12}
\langle f|\phi\rangle := \frac1{\sqrt{2\pi}} \sum_{m=-\infty}^\infty a_m^*\, e^{im\phi} \;\;\; \Longrightarrow\;\;\; \langle m|\phi\rangle = \frac1{\sqrt{2\pi}} e^{im\phi} \,.
\end{equation}
In order to prove that $|\phi\rangle$ is continuous as an antilinear functional on $\Phi$, we use the Cauchy-Schwarz inequality as follows:
\begin{equation}\begin{array}{lll}\label{13}
|\langle f|\phi\rangle| &\le &\ds \sum_{m=-\infty}^\infty |a_m| = \sum_{m=-\infty}^\infty \frac{|a_m| \,|m+i|}{|m+i|}\\[0.4cm]
 &\le&\ds \sqrt{\sum_{m=-\infty}^\infty \frac 1{|m+i|^2}}\; \sqrt{\sum_{m=-\infty}^\infty |a_m|^2\,|m+i|^2} =K\, \big|\big|\,|f\rangle\big|\big|_1\,,
\end{array}\end{equation}
where the meaning of the constant $K$ is obvious. 
Then, continuity follows from \eqref{4} and, hence, $|\phi\rangle\in\Phi^\times$. Then, let us write $\langle\phi|f\rangle:= \langle f|\phi\rangle^*$. It becomes obvious that $\langle\phi|$ is a continuous {\it linear} functional on $\Phi$. Note that
\begin{eqnarray}\label{14}
U|f\rangle =\sum_{m=-\infty}^\infty a_m U\,|m\rangle = \sum_{m=-\infty}^\infty a_m f_m(\phi) = \sum_{m=-\infty}^\infty a_m \frac 1{\sqrt{2\pi}}\, e^{-im\phi} =\langle\phi|f\rangle=f(\phi)\,.
\end{eqnarray}

Let us consider two arbitrary vectors $|f\rangle ,\, |g\rangle \in\Phi$ and their corresponding images in $\Psi$ by $U$: $f(\phi)=U\,|f\rangle$, $g(\phi)=U\,|g\rangle$, respectively. Since $U$ is unitary, it preserves scalar products, so that
\begin{equation}\label{15}
\langle g|f\rangle = \int_0^{2\pi} g^*(\phi)\,f(\phi)\,d\phi = \int_0^{2\pi} \langle g|\phi\rangle\langle \phi|f\rangle\,d\phi\,.
\end{equation}
Omiting the arbitrary $|f\rangle, \,|g\rangle \in\Phi$, so as to obtain a relation of the type
\begin{equation}\label{16}
\mathbb I =\int_0^{2\pi} |\phi\rangle\langle\phi|\,d\phi\,.
\end{equation}
Then, observe that
\begin{equation}\label{17}
|f\rangle =\mathbb I\,|f\rangle =\int_0^{2\pi} |\phi\rangle\langle\phi|f\rangle\,d\phi = \int_0^{2\pi} f(\phi)\,|\phi\rangle\,d\phi\,.
\end{equation}
Now, let us compare \eqref{11} with \eqref{17}. While \eqref{11} is a span of any vector $|f\rangle\in\Phi$ in terms of a discrete basis, \eqref{17} is a span of the same vector in terms of a {\it continuous} basis. Both bases belong to the dual space $\Phi^\times$, although the discrete basis is in both $\Phi$ and $\Phi^\times$ and the continuous basis only in $\Phi^\times$. The identity $\mathbb I$ is obviously the canonical injection from $\Phi$ into $\Phi^\times$. It is interesting that it may be inserted in the formal product $\langle\phi|f\rangle$, which is
\begin{equation}\label{18}
f(\phi)=\langle\phi|f\rangle =\int_0^{2\pi} \langle\phi|\phi'\rangle \langle\phi'|f\rangle\,d\phi'\,,
\end{equation}
so that 
\begin{equation}\label{19}
\langle\phi|\phi'\rangle =\delta(\phi-\phi')\,.
\end{equation}

Discrete and continuous bases have clear analogies. Since the basis $\{|m\rangle\}$ is an orthonormal basis in $\mathcal H$, it satisfies the following completeness relation:
\begin{equation}\label{20}
\sum_{m=-\infty}^\infty |m\rangle\langle m|=I\,,
\end{equation}
where $I$ is the identity operator on both $\mathcal H$ and $\Phi$, so that it is somehow different to the identity $\mathbb I$ \eqref{16}. The vectors $|m\rangle$ are in $\Phi$, so that they admit an expansion in terms of the continuous basis as in \eqref{17}:
\begin{equation}\label{21}
|m\rangle = \mathbb I\,|m\rangle = \int_0^{2\pi} |\phi\rangle \langle\phi|m\rangle\,d\phi = \frac1{\sqrt{2\pi}} \int_0^{2\pi} e^{-im\phi}\,|\phi\rangle\,d\phi\,.
\end{equation}
We have two identities $\mathbb I$ \eqref{16} and $I$ \eqref{20}  and both are quite different. First of all, the definitions of both identities are dissimilar. Furthermore,  $\mathbb I$ cannot be extended to an identity on $\Phi^\times$, since operations like $\langle\phi|F\rangle$ for any $F\in\Phi^\times$ cannot be defined in general. As happens with the product of distributions, only some of these brackets are allowed. For example, if $|F\rangle=|\phi'\rangle$, for $\phi'$ fixed in $[0,2\pi)$. Then, clearly 
\begin{equation}\label{20a}
\mathbb I \,|\phi'\rangle = \int_0^{2\pi} |\phi\rangle\langle\phi|\phi'\rangle\,d\phi= \int_0^{2\pi} |\phi\rangle\,\delta(\phi-\phi')\,d\phi=|\phi'\rangle\,. 
\end{equation}
On the other hand, $I$ \eqref{20} can indeed be extended to the whole $\Phi^\times$. Let us write formally for any $g\in\Phi$ and any $F\in\Phi^\times$,
\begin{equation}\label{22}
\langle g|F\rangle = \sum_{m=-\infty}^\infty \langle g|m\rangle\langle m|F\rangle\,.
\end{equation}
First of all, observe that both $\langle g|m\rangle$ and $\langle m|F\rangle$ are well defined. The question is to know whether the sum in the r.h.s. of \eqref{22} converges. To show that this is indeed the case, we need the following result:
\medskip

{\bf Lemma 1.-}{\it
For any $F\in\Phi^\times$,  there exists a constant $C>0$ and a natural $p$, such that $|\langle m|F\rangle|\le C\,|m+i|^p$.
}\smallskip

{\bf Proof.-}
It is just a mimic of the proof of Theorem V.14 in \cite{RSI}, page 143. 
\hfill$\blacksquare$\medskip

After Lemma 1, we may show the absolute convergence of the series in \eqref{22}. For that recall that $g=\sum_{m=-\infty}^\infty \langle m|g\rangle\,|m\rangle$ and that $g\in\Phi$. Then,
\begin{equation*}\begin{array}{lll}\label{23}\displaystyle
 \sum_{m=-\infty}^\infty \big|\langle g|m\rangle\big|\cdot\big|\langle m|F\rangle\big| &\le &\ds C\sum_{m=-\infty}^\infty 
 \big|\langle g|m\rangle\big|\cdot\big| m+i\big|^p = 
 C \sum_{m=-\infty}^\infty \frac{\big|\langle g|m\rangle\big|}{\big| m+i \big|^p}\,\big| m+i\big|^{2p} \\[0.4cm]  
 &\le & \ds C\, \sqrt{\sum_{m=-\infty}^\infty \big|\langle g|m\rangle\big|^2\cdot\big| m+i\big|^{4p}} \times
 \sqrt{\sum_{m=-\infty}^\infty \frac 1{\big| m+i \big|^{2p}}} =K\,\big|\big| g\big|\big|_{2p}\,,
\end{array}\end{equation*}
where $K=C$ times the second square root, which obviously converges. This shows the absolute convergence of \eqref{22}. In consequence, the formal procedure of inserting the identity in \eqref{20} to $\langle g|f\rangle$ as in \eqref{22} is rigorously correct. Thus, we see that there exists a substantial difference between the identities \eqref{16} and \eqref{20}. In addition, \eqref{20} gives a span of $|\phi\rangle$ in terms of the discrete basis $\{|m\rangle\}$ as follows:
\begin{equation}\label{24}
I|\phi\rangle =|\phi\rangle = \sum_{m=-\infty}^\infty |m\rangle\langle m|\phi\rangle = \frac1{\sqrt{2\pi}} \sum_{m=-\infty}^\infty e^{im\phi}\,|m\rangle\,.
\end{equation}
Compare \eqref{24} with the converse relation given by \eqref{21}. It is easy to prove that the series in the r.h.s. of \eqref{24} converges in the weak topology on $\Phi^\times$. 
\subsection{About representations of $SO(2)$}

We define the regular representation of $SO(2)$, $\mathcal R(\theta)$, on $L^2[0,2\pi)$ as 
\begin{equation}\label{25}
[\mathcal R(\theta)\,f](\phi):= f(\phi-\theta)\,,\quad {\rm mod}\;2\pi\,,\qquad \forall\,f\in L^2[0,2\pi)\,,
\;\;\forall \theta\in[0,2\pi)\,.
\end{equation}
 This induces an equivalent representation, $R(\theta)$, supported on $\mathcal H$ by means of the unitary mapping $U$ as
\begin{equation}\label{26}
R(\theta):= U^{-1}\mathcal R(\theta)U\,.
\end{equation}
These representations preserve the RHS structure due to  the following result:\medskip

{\bf Lemma 2.- }{\it
For any $\theta\in[0,2\pi)$, $R(\theta)$ is a bicontinuous bijection on $\Phi$.
}\smallskip

{\bf Proof.- }
Let $|f\rangle\in\Phi$ with $\ds f=\sum_{m=-\infty}^\infty a_m\,|m\rangle$. Then,
\begin{equation*}\begin{array}{lll}\label{27}
\ds R(\theta) \sum_{m = -\infty}^\infty a_m\,|m\rangle &=&\ds  U^{-1} \sum_{m=-\infty}^\infty a_m\,U\, R(\theta)\,U^{-1}\,U\,|m\rangle = U^{-1} \sum_{m=-\infty}^\infty a_m\,\mathcal R(\theta)\,\frac 1{\sqrt{2\pi}}\,e^{-im\phi} \\[0.4cm] 
&=&\ds U^{-1} \sum_{m=-\infty}^\infty a_m\,\frac 1{\sqrt{2\pi}} \,e^{-im(\phi-\theta)} = \sum_{m=-\infty}^\infty a_m\,e^{im\theta}\,|m\rangle \in \Phi\,.
\end{array}\end{equation*}
Hence, $R(\theta)\Phi\subset \Phi$. Since $R^{-1}(\theta)=R(-\theta)$, we have that $R(-\theta)\Phi\subset \Phi$, so that $\Phi\subset R(\theta)\Phi$ and, consequently, $R(\theta)\Phi =  \Phi$. 

The continuity of $R(\theta)$ on $\Phi$ is trivial for any $\theta\in[0,2\pi)$ and, hence, its inverse is also continuous. 
\hfill$\blacksquare$\medskip

This result has some immediate consequences, such as i.) $R(\theta)$ can be extended to a continuous bijection on $\Phi^\times$, as a consequence of the duality formula \eqref{8}; and ii.) $\mathcal R(\theta)$ is a bicontinuous bijection on $\Psi$ and also on $\Psi^\times$. A simple consequence of i.) is the following: since $f(\phi)=\langle\phi|f\rangle$ for all $f(\phi)\in\Psi$, we have that
\begin{equation}\label{28}
\langle R(\theta)\phi|f\rangle = [R(-\theta)\,f](\phi) =f(\phi+\theta)= \langle \phi+\theta|f\rangle\,,
\end{equation}
so that for any arbitrarily fixed $\theta\in[0,2\pi)$,
\begin{equation}\label{29}
\langle R(\theta)\phi| =\langle \phi|R(\theta) =\langle \phi+\theta| \Longleftrightarrow R(\theta)|\phi\rangle=|\theta+\phi\rangle\,, \qquad {\rm mod}\,2\pi.
\end{equation}

In addition to the regular representation, there exists one unitary irreducible representation, UIR in the sequel,  on $L^2[0,2\pi)$ for each value of $m\in\mathbb Z$ given by $\mathcal U_m(\phi):= e^{-im\phi}$. This induces a UIR on $\mathcal H$ given by $U_m(\theta)=U^{-1}\mathcal U(\phi) \,U= e^{-iJ\phi}$, where $J$ is the self-adjoint generator of all these representations. We know that for all $m\in\mathbb Z$, we have that
\begin{equation}\label{30}
J|m\rangle =m|m\rangle\,.
\end{equation}
Obviously, $J$ cannot be extended to a bounded operator on $\mathcal H$.\medskip

{\bf Proposition 1.-}{\it 
The self-adjoint operator $J$ is a well defined continuous linear operator on $\Phi$. 
}\smallskip

{\bf Proof.-} 
We define the action of $J$ on any $\ds |f\rangle=\sum_{m=-\infty}^\infty a_m\,|m\rangle \in\Phi$ as
\begin{equation}\label{31}
J|f\rangle:= \sum_{m=-\infty}^\infty a_m\,m\,|m\rangle\,.
\end{equation}
Then, for $p=0,1,2,\dots$, we have that
\begin{equation}\label{32}
\big|\big|J|f\rangle\big|\big|_p^2 = \sum_{m=-\infty}^\infty \big| a_m\big|^2 \,m^2\,\big |m+i\big|^{2p} \le \sum_{m=-\infty}^\infty \big|a_m\big|^2\cdot\big|m+i\big|^{2p+2}=\big|\big|\,|f\rangle\big|\big|_{p+1}^2\,,
\end{equation}
which shows that for any $|f\rangle\in\Phi$, $J|f\rangle$ is a well defined vector on $\Phi$. This also shows the inequality valid for any $p=0,1,2,\dots$ and all $|f\rangle\in\Phi$,
\begin{equation}\label{33}
\big|\big| J|f\rangle\big|\big|_p \le \big|\big|\,|f\rangle\big|\big|_{p+1}\,,
\end{equation}
which proves the continuity of $J$ on $\Phi$, after \eqref{5}. 
\hfill$\blacksquare$\medskip

All these properties show that $J$ may be extended to a weakly continuous linear operator on $\Phi^\times$. In order to determine its action on the functionals $|\phi\rangle$, let us consider the following derivation valid for all $f(\phi)\in\Psi$:
\begin{equation}\label{34}
i\frac{d}{d\phi}\,f(\phi) = i\frac{d}{d\phi} \sum_{m=-\infty}^\infty a_m\,e^{-im\phi} := \sum_{m=-\infty}^\infty a_m\,m\,e^{-im\phi}\,.
\end{equation}
It is a very simple exercise to show that this derivation is a well defined continuous linear operator on $\Psi$. Then, we define $D_\phi$ as
\begin{equation}\label{35}
iD_\phi:= U^{-1} \,i\frac{d}{d\phi}\,U\,.
\end{equation}
The operator $iD_\phi$ is continuous and linear on $\Phi$. Moreover, it is symmetric on $\Phi$, so that it may be extended to a weakly continuous linear operator on $\Phi^\times$. In addition:
\begin{equation}\label{36}
J|\phi\rangle =\sum_{m=-\infty}^\infty e^{-im\phi}\,J|m\rangle =\sum_{m=-\infty}^\infty e^{-im\phi}\, m\,|m\rangle = i D_\phi\,|\phi\rangle\,.
\end{equation}
This derivation is somehow unnecessary as we know from \eqref{31}, \eqref{34} and \eqref{36} that $J=iD_\phi$. Here, we close the discussion on $SO(2)$.

\section{$SU(2)$ and Associated Laguerre Functions}\label{su2associatedlaguerre}

In the previous section, we have studied the relations between the Lie group $SO(2)$, the special functions $f_m(\phi)=\frac1{\sqrt{2\pi}}\,e^{-im\phi}$, for $m\in\mathbb Z$. We have constructed a couple of RHS, one based in the use of these functions, the other being an abstract RHS unitarily equivalent to the former. In the sequel, we are going to extend a similar formalism using instead the group $SU(2)$ and the associated Laguerre functions \cite{CO17,CGO18}. 

The associate Laguerre functions \cite{SZE,AS72,OLBC}, $L_n^{(\alpha)}(x)$ are functions depending for $n=0,1,2,\dots$ on the non-negative real variable $x\in[0,\infty)$ and a fixed complex parameter $\alpha$, which satisfy the following differential equation:
\begin{equation}\label{37}
\left[x\,\frac{d^2}{dx^2} +(1+\alpha-x)\,\frac{d}{dx}+n  \right] L_n^{(\alpha)}(x) =0\,,\qquad n=0,1,2,\dots\,.
\end{equation}
Note that for $\alpha=0$, we obtain the Laguerre polynomials. In this presentation and for reasons to be clarified later, we are interested in those associated Laguerre functions such that $\alpha$ be an integer number, $\alpha\in\mathbb Z$.
\subsection{Associated Laguerre Functions}

 It is also useful to introduce a set of alternative variables, such as $j:= n+\alpha/2$ and $m:=-\alpha/2$ with $|m|\le j$ and $j-m\in\mathbb N$, $\mathbb N$ being the set of non-negative integers. Observe that $j$ is either positive integer or positive semi-integer, i.e. $n \in \mathbb N$, $\alpha \in \mathbb Z$ and $\alpha \ge -n$. Then, we define the following sequence of functions:
\begin{equation}\label{38}
\mathcal L_j^m(x):= \sqrt{\frac{(j+m)!}{(j-m)!}}\, x^{-m}\,e^{-x/2}\, L_{j+m}^{(-2m)} (x)\,.
\end{equation}
These functions are symmetric with respect to to the exchange $m \leftrightarrow -m$. In addition, they satisfy the following orthonormality and completeness relations:
\begin{equation}\label{39}
\int_0^\infty \mathcal L_j^m(x)\,\mathcal L_{j'}^m(x)\,dx = \delta_{jj'} \,,\qquad \sum_{j=|m|}^\infty \mathcal L_j^m(x)\,\mathcal L_j^m(x') =\delta(x-x')\,.
\end{equation}
It is also well known that, for a fixed value of $m$, the functions $\{\mathcal L_j^m(x)\}_{j=|m|}^\infty$ form an orthonormal basis for $L^2(\mathbb R^+)$, $\mathbb R^+:=[0,\infty)$. 

We may rewrite the differential equation  \eqref{37} in terms of the functions $\mathcal L_j^m(x)$ as
\begin{equation}\label{40}
\left[ X\,D_x^2 +D_x -\frac 1X\,M^2 -\frac X4 +J+\frac 12 \right] \mathcal L_j^m(x) =0\,,
\end{equation}
where
\begin{equation}\begin{array}{ll}\label{41}
X\,\mathcal L_j^m(x) := x\,\mathcal L_j^m(x) \,, \qquad & D_x\,
\mathcal L_j^m(x):= \frac{d}{dx}\,\mathcal L_j^m(x)\,,\\[0.4cm]
 J\,\mathcal L_j^m(x) :=j\,\mathcal L_j^m(x)\,,\qquad & M\,\mathcal L_j^m(x):= m\,\mathcal L_j^m(x)\,.
\end{array}\end{equation}
The operators in \eqref{41} can be extended by linearity and closeness  to domains dense in $L^2(\mathbb R^+)$. Next, we formally define the following linear operators:
\begin{equation}\begin{array}{lll} \label{42}
K_\pm &:=& \mp 2D_x \left(M\pm\frac 12 \right) +\frac 2X \left(M\pm\frac 12 \right) - \left( J+\frac 12 \right)\, 
\\[0.4cm]
K_3 &:=& M \,,
\end{array}\end{equation}
which give the following relations:
\begin{equation}\begin{array}{lll} \label{45}
K_\pm \,\mathcal L_j^m(x) &:=& \sqrt{(j\mp m)(j\pm m+1)}\,\mathcal L_j^{m\pm 1}(x)\,  \\[0.4cm]
K_3 \, \mathcal L_j^m(x) &:=& m\, \mathcal L_j^m(x) \,.
\end{array}\end{equation}
On the subspace spanned by linear combinations of the functions $\mathcal L_j^m(x)$, this gives the following commutation relations:
\begin{equation}\label{48}
[K_+,K_-]=2K_3\,,\qquad [K_3,K_\pm]=\pm K_\pm\,,
\end{equation}
which are the commutation relations for the generators of the Lie algebra $su(2)$. For each fixed value of $j$ integer or half-integer
and $-j\le m\le j$, the space of the linear combinations of the functions $\mathcal L_j^m(x)$ support a $2j+1$ dimensional representation of $SU(2)$. 
\subsection{Associated Laguerre functions on the plane}

In RHS the number of variables is equal to the number of parameters
because the properties of $\Phi$ and $\Phi^\times$. In subsection 3.1
we discussed a RHS based on
one parameter $m$ and one continuous variable $x$. An alternative is to introduce a
new continuous variable $\phi$ and construct a RHS with two parameters $j$ and $m$ and
two variables, the old one $x$ and this new one $\phi$. This point will be
discussed in general in Section \ref{conclusions}.

Then, we introduce an angular variable $\phi\in[-\pi,\pi]$ and the new functions:
\begin{equation}\label{49}
\mathcal Z_j^m(r,\phi):= e^{im\phi}\,\mathcal L_j^m(r^2)\,.
\end{equation}
These functions satisfy the property $\mathcal Z_j^m(r,\phi+2\pi) = (-1)^{2j} \mathcal Z_j^m(r,\phi)$. After \eqref{40} and the change of variable $x\to r^2$, we obtain the following differential equation for $\mathcal Z_j^m(r,\phi)$:
\begin{equation}\label{50}
\left[\frac{d^2}{dr^2} + \frac 1r\,\frac{d}{dr} -\frac{4m^2}{r} -r^2 +4\left(j+\frac 12 \right) \right] \mathcal Z_j^m(r,\phi)=0\,.
\end{equation}

It is not difficult to obtain the orthonormality and completeness relations for the functions $\mathcal Z_j^m(r,\phi)$, which are
\begin{equation}\begin{array}{l}\label{51}
\ds \frac 1\pi \int_{-\pi}^\pi d\phi \int_0^\infty r\,dr\, [\mathcal Z_j^m(r,\phi)]^* \,\mathcal Z_{j'}^{m'}(r,\phi) =\delta_{jj'}\,\delta_{mm'}\,,\\[0.4cm]
\ds \sum_{j,m} [\mathcal Z_j^m(r,\phi)]^* \, \mathcal Z_j^m(r',\phi') = \frac \pi r\,\delta(r-r')\,\delta(\phi-\phi')\,.
\end{array}\end{equation}
This shows that the set of functions $\{\mathcal Z_j^m(r,\phi)\}$ forms a basis of $L^2(\mathbb R^2,d\mu)$ with $d\mu(r,\phi):=r\,dr\,d\phi/\pi$. Observe the similitude with the set of spherical harmonics $\{Y_j^m(\theta,\phi)\}$, which forms a basis of the Hilbert space $L^2(S^2,d\Omega)$. 

Let $\mathcal H$ be an abstract infinite dimensional separable Hilbert space and $U$ a unitary mapping from $L^2(\mathbb R^2)$ onto $\mathcal H$, $U:L^2(\mathbb R^2)\longmapsto \mathcal H$. An orthonormal basis $\{|j,m\rangle\}$ in $\mathcal H$ is given by $|j,m\rangle: = U \mathcal Z_j^m(r,\phi)$, so that  $\{|j,m\rangle\}$ satisfy the conditions of orthonormality and completeness:
\begin{equation}\label{53}
\langle j,m|j',m'\rangle=\delta_{jj'}\,\delta_{mm'}\,,\qquad \sum_{j_{min}}^\infty \sum_{m=-j}^j |j,m\rangle\langle j,m|=\mathcal I\,,
\end{equation}
where $j_{min}=0$ for integer spins and $j_{min}={1/2}$ for half-integer
spins. 

After the two last equations in \eqref{41}, we define
\begin{equation}\label{54}
\widetilde J:= UJU^{-1}\,, \qquad \widetilde M= UMU^{-1}\,,
\end{equation}
so that
\begin{equation}\label{55}
\widetilde J|j,m\rangle=j|j,m\rangle\,,\qquad \widetilde M|j,m\rangle =m|j,m\rangle\,,
\end{equation}
where $j$ is a non-negative integer
or half-integer and $m=-j,-j+1,\dots,j-1,j$. 

Let us proceed with the definitions of some new objects. First of all, the operators $J_\pm$ and $~J_3$ on $L^2(\mathbb R^2)$, which are

\begin{equation}\label{56}
J_\pm:= e^{\pm i\phi}K_\pm\,,\qquad J_3:= K_3\,.
\end{equation}
The operators defined in \eqref{56} act on the functions $\mathcal Z_j^m(r,\phi)$ exactly as $K_\pm$ and $K_3$ on $\mathcal L_j^m(x)$, expressions given in \eqref{45}. Also, we define the corresponding operators on $\mathcal H$ as 
\begin{equation}\label{57}
\widetilde J_\pm := U\,J_\pm\, U^{-1}\,,\qquad \widetilde J_3:= U\, J_3 \, U^{-1}\,,
\end{equation}
so that 
\begin{equation}\begin{array}{l}\label{58}
\ds\widetilde J_\pm \,|j,m\rangle = \sqrt{(j\mp m)(j\pm m+1)}\,|j,m\pm 1\rangle \,, \\[0.4cm] 
\ds\widetilde J_3 \,|j,m\rangle = m\,|j,m\rangle\,.
\end{array}\end{equation}
\subsection{Rigged Hilbert spaces associated to $L^2(\mathcal R^2)$}

On $\mathcal H$, we have already defined a pair of the discrete basis $\{|j,m\rangle\}$, one for integer values of $j$ and the other for half-integer values of $j$. In order to define continuous bases, we have to construct a suitable pair of RHS. Let us consider the space of all $|f\rangle\in\mathcal H$,
\begin{equation}\label{59}
|f\rangle =\sum_{j_{min}}^\infty \sum_{m=-j}^j a_{j,m}\,|j,m\rangle\qquad {\rm with} \quad \sum_{j_{min}}^\infty \sum_{m=-j}^j |a_{j,m}|^2<\infty\,,
\end{equation}
where we have taken one of the choices for $j$, either integer or half-integer, such that they satisfy the following property:
\begin{equation}\label{60}
\big|\big|\,|f\rangle\big|\big|^2_p:=\sum_{j_{min}}^\infty\, \sum_{m=-j}^j \big|a_{j,m}\big|^2 \left(2^{3|m|}(j+|m|+1)!\right)^{2p}<\infty\,, \qquad p=0,1,2,\dots\,,
\end{equation} 
where, again, we may use either the basis with $j$ integer or with $j$ half-integer. We call $\Phi_I$ and $\Phi_H$ the resulting spaces, where the indices $I$ and $H$ mean ``integer'' and ``half-integer'', respectively. This spaces are rather small. Nevertheless, they are still dense in $\mathcal H$, since they contain the orthonormal basis $\{|j,m\rangle\}$. We need this kind of topology in order to guarantee the continuity of the elements of the continuous basis, as shall see. Norms $\big|\big|-\big|\big|_p$ endow both $\Phi_I$ and $\Phi_H$ of a structure of metrizable locally convex space and give a pair of unitarily equivalent RHS
\begin{equation}\label{61}
\Phi_I\subset \mathcal H \subset \Phi_I^\times\,,\qquad \Phi_H\subset \mathcal H \subset \Phi_H^\times\,.
\end{equation}
On these structures, it makes sense the existence of continuous bases, $\{|r,\phi\rangle\}$, as we can show right away. For each pair of values of $r$ and $\phi$, we define the following anti-linear mapping $|r,\phi\rangle$ as follows. Let $f(r,\phi):= U^{-1}\,|f\rangle$, $|f\rangle\in\mathcal H$ so that
\begin{equation}\label{62}
f(r,\phi)=\sum_{j_{min}}^\infty \sum_{m=-j}^j a_{j,m}\,\mathcal Z_j^m(r,\phi)\,.
\end{equation}
Then, define
\begin{equation}\label{63}
\langle f|r,\phi\rangle:= \sum_{j_{min}}^\infty \sum_{m=-j}^j a_{j,m}^* [\mathcal Z_j^m(r,\phi)]^*\,.
\end{equation}
Note that for $|f\rangle=|j,m\rangle$, we have that
\begin{equation}\label{64}
\langle j,m|r,\phi\rangle = [\mathcal Z_j^m(r,\phi)]^* \quad {\rm or} \quad \langle r,\phi|j,m\rangle =\langle j,m|r,\phi\rangle^* =\mathcal Z_j^m(r,\phi)\,.
\end{equation} 
As in the previous cases, we may define $\Psi_{I,H}:= U^{-1} \Phi_{I,H}$, so as to define two new RHS, which are unitarily equivalent to the \eqref{61}. These are 
\begin{equation}\label{65}
\Psi_{I,H} \subset L^2(\mathbb R^2) \subset \Psi_{I,H}\,.
\end{equation}
Then, $f(r,\phi)$ as in \eqref{62} is in $\Psi_I$ or in $\Psi_H$, if and only if the coefficients $a_{j,m}$ satisfy the relations \eqref{60}.
The kets $|r,\phi\rangle$, which are obviously linear on $\Phi_I$ and $\Phi_H$, are also continuous under the topologies induced by the norms $||-||_p$. This is a consequence of the next two results.\medskip

{\bf Lemma 3.- }{\it 
The functions $\mathcal Z_j^m(r,\phi)$ have the following upper bound:
\begin{equation}\label{66}
\big|\mathcal Z_j^m(r,\phi)\big| \le 2^{3|m|} \,\frac{(j!)^2 [(j+|m|)!]^{1/2}}{|m|! [(j-|m|)!]^{5/2}}\,.
\end{equation}
}

{\bf Proof.-} 
To begin with, look at equation \eqref{38} and \eqref{49}. Then, we use the following inequality, which has been given in \cite{DUR}:
\begin{equation}\label{67}
|x^k L_n^{(\alpha)}(x) e^{-x/2}| \le 2^{\min(\alpha,k)}\,2^k (n+1)^{(k)} \left( \begin{array}{c} n+\max(\alpha-k,0) \\ n \end{array} \right)\,.
\end{equation}
Here, $k$ and $n$ are natural numbers, $\alpha\ge 0$, $x\ge 0$ and $(n+1)^{(k)} := (n+1)(n+2)\dots (n+k)$ is the Pochhammer symbol.

We have to consider the cases $m<0$ and $m\ge 0$, as well as the condition $\alpha\ge 0$, which is necessary for the validity of inequality \eqref{4}. All these two conditions are really only one since $m=-\alpha/2$ and the functions $\mathcal L_j^m(x)$ satify  the following symmetry relation:
\begin{equation}\label{68}
\mathcal L_j^m(x)=(-1)^{2j}\,\mathcal L_j^{-m}(x)\,.
\end{equation} 
Then, we discuss $m<0$. Here, we write $-m$ with $m\in\mathbb N$ instead. Take \eqref{38}, where we replace $m$ by $-m$ and use \eqref{67}. First, we have
\begin{equation}\label{69}
\big| x^k L_n^{(\alpha)}(x) e^{-x/2}\big| \le 2^{3m} (j-m+1)^{(m)} \left(\begin{array}{c} j \\ j-m \end{array}  \right)
\end{equation}
Then, complete $\mathcal L^{-m}_j(x)$ so as to obtain
\begin{equation}\label{70}
\big| \mathcal L_j^{-m}(x)\big| \le 2^{3m}\,\frac{(j!)^2 [(j+m)!]^{1/2}}{m! [(j-m)!]^{5/2}}\,.
\end{equation}
This result, along \eqref{68} and \eqref{49} gives \eqref{66}. 
\hfill $\blacksquare$
\medskip

{\bf Theorem 2.-}{\it 
Each of the kets $|r,\phi\rangle$ is a continuous anti-linear functional in both $\Phi_I$ and $\Phi_H$. 
}\smallskip

{\bf Proof.-}
It is a consequence of the previous lemma 3. From \eqref{63} and \eqref{66}, we have the following inequalities, the first one in the second row being the Cauchy-Schwarz inequality,
\begin{equation}\begin{array}{l}\label{71}
\ds \big|\langle f|r,\phi\rangle\big|= \sum_{j_{min}}^\infty \sum_{m=-j}^j \big|a_{j,m}\big|\, \big|\mathcal Z_j^m(r,\phi)\big| \le \sum_{j_{min}}^\infty \sum_{m=-j}^j |a_{j,m}|\, 2^{3|m|} \,\frac{(j!)^2 ((j+|m|)!)^{1/2}}{|m|! ((j-|m|)!)^{5/2}} \\[0.5cm] 
\ds\qquad \qquad\le \sqrt{\sum_{j_{min}}^\infty \sum_{m=-j}^j |a_{j,m}|^2\, 2^{6|m|} \,  (j!)^2 \, (j+|m|)!  }\; \times \; \sqrt{\sum_{j_{min}}^\infty \sum_{m=-j}^j \frac{1}{(|m|!)^2\,((j-|m|)!)^5}}\,.
\end{array}\end{equation}
The second row in \eqref{71} is the product of two terms. The second one is the root of a convergent series. Let us denote this term by $C>0$. The expression under the square root in the first factor is bounded by
\begin{equation}\label{72}
\sum_{j_{min}}^\infty \sum_{m=-j}^j \big| a_{j,m}\big|^2\, \left(2^{3|m|} (j+|m|+1)!\right)^4 = \big|\big|\,|f\rangle \big|\big|^2_2\,,
\end{equation}
so that
\begin{equation}\label{73}
\big|\langle f|r,\phi\rangle\big| \le C\, \big|\big|\,|f\rangle\big|\big|_2\,,
\end{equation}
which, along the linearity of $|r,\phi\rangle$ on $\Phi_{I,H}$, proves our assertion.
\hfill $\blacksquare$
\medskip

Formal relations between discrete $\{|j,m\rangle\}$ and continuous bases $\{|r,\phi\rangle\}$ are easy to find. Let us go back to \eqref{63}. Due to the unitary relation between $L^2(\mathbb R^+)$ and $\mathcal H$, we conclude that $a_{j,m}^*=\langle f|j,m\rangle$, so that $\ds \langle f|r,\phi\rangle = \sum_{j=0}^\infty \sum_{m=-j}^j \langle f|j,m\rangle \,\mathcal Z_j^m(r,\phi)$ and, hence, omitting the arbitrary $|f\rangle\in\Phi_{I,H}$, we have that
\begin{equation}\label{74}
|r,\phi\rangle = \sum_{j_{min}}^\infty \sum_{m=-j}^j  \mathcal Z_j^m(r,\phi) \, |j,m\rangle\,.
\end{equation}
The inverse relation may be easily found taking into account the unitary mapping between $L^2(\mathbb R^+)$ and $\mathcal H$, again. In fact, being given $|f\rangle,|g\rangle\in \Phi_{I,H}$, their scalar product gives:
\begin{equation}\label{75}
\langle f|g\rangle= \int_0^{2\pi} d\phi \int_0^\infty r\, dr\, \langle f|r,\phi\rangle\langle r,\phi|g\rangle \,.
\end{equation}
Then, if we choose $|g\rangle=|j,m\rangle$ and omit the arbitrary $|f\rangle\in\Phi_{I,H}$, we have the converse relation to \eqref{74} as
\begin{equation}\label{75}
|j,m\rangle = \int_0^{2\pi} d\phi \int_0^\infty r\,dr\, |r,\phi\rangle\langle r,\phi|j,m\rangle = \int_0^{2\pi} d\phi \int_0^\infty r\,dr\, \mathcal Z_j^m(r,\phi)\, |r,\phi\rangle\,.
\end{equation}
Although this is implicit in the above expressions, it could be interesting to write the explicit spans of any $|f\rangle\in\Phi_{I,H}$ in terms of the discrete and continuous basis. These are
\begin{equation}\label{76}
|f\rangle =\sum_{j_{min}}^\infty \sum_{m=-j}^j \langle j,m|f\rangle\,|j,m\rangle\,,
\end{equation}
and
\begin{equation}\label{77}
|f\rangle = \int_0^{2\pi} \int_0^\infty r\,dr\, f(r,\phi)\,|r,\phi\rangle\,.
\end{equation}

The continuity of the linear operators $\widetilde J$, $\widetilde M$, $\widetilde J_\pm$ and $\widetilde J_3$ is rather obvious. For instance, for any $|f\rangle\in\Phi_{I,H}$, we define 
\begin{eqnarray}\label{78}
\widetilde J |f\rangle = \sum_{j_{min}}^\infty \sum_{m=-j}^j a_{j,m} \,j\,|j,m\rangle\,,
\end{eqnarray}
so that, for any $p=0,1,2,\dots$
\begin{equation}\begin{array}{lll}\label{79}
\big|\big|\widetilde J\,|f\rangle\big|\big|^2_p &:=&\ds \sum_{j_{min}}^\infty \sum_{m=-j}^j |a_{j,m}|^2\, j^2 [2^{3|m|}(j+|m|+1)!]^{2p}
 \\[0.4cm] 
&\le&\ds \sum_{j_{min}}^\infty \sum_{m=-j}^j |a_{j,m}|^2 \left(2^{3|m|}(j+|m|+1)!\right)^{2(p+1)} = \big|\big|\,|f\rangle\big|\big|^2_{p+1}\,.
\end{array}\end{equation}
This relation proves both, that $\widetilde J\,|f\rangle $ is in either $\Phi_{I,H}$ and the continuity of $\widetilde J$ in both spaces. Similar results can be obtained for the other operators: $\widetilde M$, $\widetilde J_\pm$ and $\widetilde J_3$ .  

As a matter of fact, the topology \eqref{60} is too strong, if we just wanted to provide RHS for which the above operators be continuous. Take for instance the spaces of all $|f\rangle\in\mathcal H$ such that
\begin{equation}\label{80}
q(|f\rangle)^2=\sum_{j_{min}}^\infty \sum_{m=-j}^j \big|a_{j,m}\big|^2\left(j+|m|+1\right)^{2p}\,,\qquad p=0,1,2\dots\,.
\end{equation}
One of the spaces, $\Xi_I$, holds for $j$ integer and the other, $\Xi_H$, holds for $j$ half-integer. The above operators reduce both spaces $\Xi_{I,H}$ and are continuous on them. The proof is essentially identical as in the previous case. Thus, we have two sequences of rigged Hilbert spaces one for $j$ integer, labelled by $I$, and the other for $j$ half-integer, labelled by $H$, where all the inclusions are continuous:
\begin{equation}\label{81}
\Phi_{I,H} \subset \Xi_{I,H}\subset \mathcal H \subset \Xi_{I,H}^\times \subset \Phi_{I,H}^\times\,.
\end{equation}

While the operators $\widetilde J$, $\widetilde M$, $\widetilde J_\pm$ and $\widetilde J_3$ are continuous on $\Phi_{I,H}$ and $\Xi_{I,H}$, we have introduce the topology \eqref{60} just to make sure of the continuity of the functionals $\{|r,\phi\rangle\}$. All these operators can be continuously extended to the duals. Note that $\widetilde J$, $\widetilde M$ and $\widetilde J_3$ are symmetric, although $\widetilde J_\pm$ are formal adjoint of each other. 


\section{Weyl-Heisenberg group and  Hermite functions}\label{weylheisengerghermite}

Possibly, the better studied and the most widely used of the special functions are the Hermite functions. When properly normalized, the Hermite functions form an orthonormal discrete basis for $L^2(\mathbb R)$ and have the form
\begin{equation}\label{83}
\psi_n(x):= \frac{e^{-x^2/2}}{\displaystyle \sqrt{2^n n! \sqrt\pi}}\,H_n(x)\,,
\end{equation}
where $H_n(x)$ are the Hermite polynomials  \cite{SZE,AS72,OLBC}. 

\subsection{Continuous and discrete bases and RHS}

In quantum mechanics for one-dimensional systems \cite{CT91}, one often uses a pair of continuous bases: the continuous bases in the coordinate and momentum representation, denoted as $\{|x\rangle\}$ and $\{|p\rangle\}$ respectively,  with $x,p\in\mathbb R$. Kets $|x\rangle$ and $|p\rangle$ are eigenkets of the position operator $Q$ and momentum operator $P$, respectively: $Q|x\rangle =x|x\rangle$ and $P|p\rangle=p|p\rangle$. In order to define these objects and the continuous bases they produce, we need RHS \cite{BOHM2}. 
Then, the ingredients in our construction are the following:
\begin{itemize}

\item{The Schwartz space $S$ of all complex indefinitely differentiable functions of the real variable $x\in\mathbb R$, such as they and all their derivatives at all orders go to zero at the infinity faster than the inverse of any polynomial. The Schwartz space $S$ is endowed with a metrizable locally convex topology \cite{RSI}. It is well known that $S$ is the first element of a RHS $S\subset L^2(\mathbb R)\subset S^\times$. Note that the Fourier transform leaves this triplet invariant.}
\medskip

\item{An abstract infinite-dim. separable Hilbert space $\mathcal H$ along a fixed, although arbitrary, unitary operator $U:\mathcal H\to  L^2(\mathbb R)$. If $\Phi:=U^{-1}S$ and we transport the locally convex topology from $S$ to $\Phi$ by $U^{-1}$, we have a second RHS $\Phi\subset\mathcal H\subset \Phi^\times$, unitarily equivalent to $S\subset L^2(\mathbb R)\subset S^\times$. }
\medskip

\item{For any $|f\rangle\in\Phi$ and any $x_0\in\mathbb R$, we define $\langle f|x_0\rangle := f(x_0)$, where $f(x):= U|f\rangle$, so that $f(x)\in S$. Analogously, for any $p_0\in\mathbb R$, we define
\begin{equation}\label{84}
\langle f|p_0\rangle:= \int_{-\infty}^\infty e^{-ixp_0} \,\langle f|x\rangle \,dx = \int_{-\infty}^\infty e^{-ixp_0} \, f(x)\,dx\,.
\end{equation}
Vectors $|x\rangle, |p\rangle\in\Phi^\times$ for any $x,p\in\mathbb R$ \cite{CGO1}.}
\medskip

\item{Define $\widetilde Q f(x):=xf(x)$ and $\widetilde Pf(x):= -if'(x)$, for all $f(x)\in S$, where the prime means derivative. Let $Q:= U^{-1}\widetilde Q U$ and $P:=U^{-1}\widetilde P U$. Then, for given $x_0\in\mathbb R$, $\langle Qf|x_0\rangle= x_0f(x_0)$, so that $\langle f|Q|x_0\rangle= x_0f(x_0)=x_0\langle f|x_0\rangle$, which implies that $Q|x_0\rangle =x_0|x_0\rangle$. We have used the same notation for $Q$ and its extension to $\Phi^\times$. Analogously, $P|p_0\rangle=p_0|p_0\rangle$, for any $p_0\in\Phi^\times$. }
\medskip

\item{Since $U$ is unitary, it preserves scalar products, so that for arbitrary $|g\rangle,|f\rangle\in\Phi$, we have
\begin{equation}\label{85}
\langle g|f\rangle =\int_{-\infty}^\infty g^*(x)f(x)\,dx =\int_{-\infty}^\infty \langle g|x\rangle\langle x|f\rangle \,dx\,,
\end{equation}
which defines the following identity:
\begin{equation}\label{86}
I:= \int_{-\infty}^\infty |x\rangle\langle x|\,dx\,,
\end{equation}
which is the canonical injection $I: \Phi \longmapsto \Phi^\times$ with $I|f\rangle\in \Phi^\times$ for any $|f\rangle\in \Phi$. Another representation of this identity is

\begin{equation}\label{87}
I=\int_{-\infty}^\infty |p\rangle\langle p|\,dp\,.
\end{equation}
This means that, for any $|f\rangle\in\Phi$, $I|f\rangle\in\Phi^\times$ can be written as
\begin{equation}\label{88}
I|f\rangle =\int_{-\infty}^\infty |x\rangle\langle x|f\rangle \,dx = \int_{-\infty}^\infty f(x)\,|x\rangle\,dx\,,
\end{equation}
and
\begin{equation}\label{89}
I|f\rangle =\int_{-\infty}^\infty |p\rangle\langle p|f\rangle\,dp = \int_{-\infty}^\infty f(p)\,|p\rangle\,dp\,,\quad {\rm with} \quad f(p)= \frac1{\sqrt{2\pi}} \int_{-\infty}^\infty e^{-ipx}\,f(x)\,dx\,.
\end{equation}
}

\item{The conclusion of the above paragraph is that either set of vectors $\{|x\rangle\}$ and $\{|p\rangle\}$ forms a {\it continuous basis} for the vectors in $\Phi$. In addition, we have a discrete basis on $\mathcal H$ defined as 
\begin{equation}\label{90}
|n\rangle := U^{-1}\left(\psi_n(x)\right)\,, \qquad n=0,1,2,\dots\,,
\end{equation}
which has the properties,
\begin{equation}\label{91}
\mathbb I=\sum_{n=0}^\infty |n\rangle\langle n|\,, \qquad \langle n|m\rangle=\delta_{n,m}\,,
\end{equation}
where $\mathbb I$ is the identity operator on $\mathcal H$. For any $|f\rangle = \sum_{n=0}^\infty a_n|n\rangle\in\Phi$, we have that
\begin{equation}\label{92}
a_n^* =\langle f|n\rangle =\int_{-\infty}^\infty \psi_n(x)\,f^*(x)\,dx = \int_{-\infty}^\infty \psi_n(x)\,\langle f|x\rangle \,dx\,,
\end{equation}
so that
\begin{equation}\label{93}
|n\rangle =\int_{-\infty}^\infty \psi_n(x)\,|x\rangle\,dx\,,\qquad n=0,1,2,\dots\,,
\end{equation}
identity that makes sense in $\Phi^\times$. Taking into account \eqref{86}, \eqref{92} and that $\langle n|x\rangle=\psi_n(x)$, we may invert formula \eqref{93}. Take an arbitrary 
$|f\rangle\in\Phi$:
\begin{equation}\begin{array}{lll}\label{94}
\ds \langle f|x\rangle &=&\ds \sum_{n=0}^\infty a_n^* \langle n|x\rangle = \sum_{n=0}^\infty \left[\int_{-\infty}^\infty \psi_n(x')\,\langle f|x'\rangle\,dx \right] \langle n|x\rangle \\[0.4cm]
& =&\ds \sum_{n=0}^\infty \left[\int_{-\infty}^\infty \langle f|x'\rangle\langle x'|n\rangle \,dx\right] \psi_n(x) = \sum_{n=0}^\infty \psi_n(x) \,\langle f|n\rangle\,,
\end{array}\end{equation}
so that, if we omit the arbitrary bra $\langle f|$, we conclude that
\begin{equation}\label{95}
|x\rangle = \sum_{n=0}^\infty \psi_n(x) \, |n\rangle\,, 
\end{equation}
identity that makes sense in $\Phi^\times$. Another property can be easily shown from \eqref{70} and $f(x)=\langle x|f\rangle\in\Phi$: 
\begin{equation}\label{96}
f(x')= \langle x'|f\rangle = \int_{-\infty}^\infty f(x) \langle x'|x\rangle\,dx \Longleftrightarrow \langle x'|x\rangle =\delta(x-x')\,.
\end{equation} 
} 

\item{Analogously, in the momentum representation, we have that
\begin{equation}\label{97}
|n\rangle = (-i)^n \int_{-\infty}^\infty \psi_n(p) \,|p\rangle\,dp\,,
\end{equation}
since the Fourier transform of $\psi_n(x)$ is $(-i)^n\psi_n(p)$ 
\begin{equation}\label{98}
|p\rangle = \sum_{n=0}^\infty (-i)^n \psi_n(p)\,|n\rangle\,,
\end{equation}
and $\langle p|p'\rangle = \delta(p-p')$. 
}

\end{itemize}

\subsection{The Weyl-Heisenberg Lie algebra}

Let us consider the following operators, defined by their action on the normalized Hermite functions \cite{CO13} :
\begin{equation}\label{99}
\widetilde{Q}\psi_n(x):= x\psi_n(x)\,,\quad \widetilde{P}\psi_n(x):=i\frac{d\psi_n(x)}{dx}\,, \quad \widetilde{N}\psi_n(x) := n\psi_n(x)\,,\quad \widetilde{\mathbb I} \psi_n(x):=\psi_n(x)\,
\end{equation}
for $n=0,1,2,\dots$. These operators can be uniquely extended to $S$, and these extensions are essentially self-adjoint  and continuous on $S$ with its own topology, so that they are extensible to weakly continuous operators on $S^\times$.  The properties of these operators are very well known. Let us name
\begin{equation}\label{100}
Q:= U\widetilde Q U^{-1}\,,\qquad P:= U\widetilde P U^{-1}\,,\qquad N:= U\widetilde N U^{-1}\,,\qquad \mathbb I:= U\widetilde {\mathbb I} U^{-1}\,,
\end{equation}
which have the same properties on $\Phi$. As usual,
\begin{equation}\label{101}
a:= \frac1{\sqrt 2}\,(Q-iP)\,,\qquad a^\dagger:= \frac 1{\sqrt 2} \, (Q+iP)\,,
\end{equation}
so that,
\begin{equation}\label{102}
a\,|n\rangle = \sqrt n\,|n-1\rangle\,,\qquad a^\dagger\,|n\rangle =\sqrt{n+1}\,|n+1\rangle\,.
\end{equation}
Obviously, $a$ and $a^\dagger$ are continuous on $\Phi$ and extended with continuity to $\Phi^\times$. The extensions are defined using the duality formula \eqref{6}. As a system of generators of the Weyl-Heisenberg Lie algebra, we may use either $\{Q,P,N,\mathbb I\}$ or $\{a, a^\dagger,N,\mathbb I\}$. Note that
\begin{equation}\label{103}
N=\frac 12 (\{a,a^\dagger \}-\mathbb I)=\frac 12 (Q^2+P^2-\mathbb I)\,,
\end{equation}
where the brackets mean anti-commutator. On $\Phi$, the Casimir operator vanishes:
\begin{equation}\label{104}
C:= \frac 12 (Q^2+P^2)-\left( N+ \frac 12 \,\mathbb I\right) \equiv 0\,.
\end{equation}

In addition, the universal enveloping algebra of the Weyl-Heisenberg group is irreducible on the RHS $\Phi\subset\mathcal H\subset \Phi^\times$.  

\section{The group SO(3,2) and the spherical harmonics}\label{so32sphericalharmonics}

Let us consider the hollow unit sphere $S^2$ in $\mathbb R^3$. Any point in $S^2$ is characterized by two angular variables $\theta$ and $\phi$, with $0\le \theta\le \pi$ and $0\le\phi<2\pi$. Let us consider the Hilbert space, $L^2(S^2,d\Omega)$, with $d\Omega:= d(\cos\theta)\,d\phi$, of Lebesgue measurable complex functions, $f(\theta,\phi)$, such that
\begin{equation}\label{105}
\int_0^{2\pi} d\phi \int_0^\pi d(\cos\theta) \,|f(\theta,\phi)|^2<\infty\,.
\end{equation}

An orthonormal basis for $L^2(S^2,d\Omega)$ is given by $\sqrt{l+1/2}\,Y_l^m(\theta,\phi) $, where $Y_l^m(\theta,\phi)$ are the spherical harmonics  \cite{SZE,AS72,OLBC}
\begin{equation}\label{106}
Y_l^m(\theta,\phi) =\sqrt{\frac{(l-m)!}{2\pi\,(l+m)!}}\,e^{im\phi}\,P_l^m(\cos\theta)\,,
\end{equation}
where $l\in\mathbb N$, the set of natural numbers, $m\in\mathbb Z$ the set of integers, with $|m|\le l$ and $P_l^m$ are the associated Legendre functions. This means, in particular, that for any $f(\theta,\phi)\in L^2(S^2,d\Omega)$
\begin{equation}\label{107}
f(\theta,\phi) = \sum_{l=0}^\infty \sum_{m=-l}^l f_{l,m}\, \sqrt{l+1/2}\,Y_l^m(\theta,\phi) \,, \quad {\rm with} \quad \sum_{l=0}^\infty \sum_{m=-l}^l |f_{l,m}|^2<\infty\,,
\end{equation}
and
\begin{equation}\label{108}
f_{l,m} = \sqrt{l+1/2} \int_0^{2\pi} d\phi \int_0^\pi d(\cos\theta) \, Y_l^m(\theta,\phi)^* \,f(\theta,\phi)\,.
\end{equation}
From the fact that the set of spherical harmonics is an orthonormal basis, we obtain the following relations:
\begin{equation}\begin{array}{rll}\label{110}
\ds \int_{S^2} d\Omega \,  Y_l^m(\theta,\phi)^*(l+1/2) Y_l^m(\theta',\phi')& =&\ds 
\delta_{l,l'}\,\delta_{m,m'}\,, 
\\[0.4cm]
\ds \sum_{l=|m|}^\infty \sum_{m=-\infty}^\infty Y_l^m(\theta,\phi)^*(l+1/2) Y_l^m(\theta',\phi') &=&\ds
\delta(\cos\theta-\cos\theta')\,\delta(\phi-\phi') \,, 
\end{array}\end{equation}
with $\delta(\cos\theta-\cos\theta') = \delta(\theta-\theta')/ \big|\sin \theta\big|$. 


\subsection{RHS associated to the spherical harmonics}

The Hilbert space $L^2(S^2,d\Omega)$ supports a representation of a UIR of the de-Sitter group $SO(3,2)$ with quadratic Casimir $\mathcal C_{so(3,2)}=-5/4$ on the spherical harmonics \cite{CO,CGO2}. The action of  the generators of the Cartan subalgebra of the Lie algebra $so(3,2)$, $L$ and $M$, is

\begin{equation}\label{111}
L\,Y_l^m(\theta,\phi) =l\,Y_l^m(\theta,\phi)\,,\qquad M\, Y_l^m(\theta,\phi) =m\,Y_l^m(\theta,\phi)\,.
\end{equation}
Once we have established this Hilbert space which supports a representation of the Anti-de-Sitter group , let us consider a unitarily equivalent abstract Hilbert space $\mathcal H\equiv U[ L^2(S^2,d\Omega)]$, where $U$ is unitary. An orthonormal basis for this space is given by the vectors $\{|l,m\rangle\}$, where for each pair $l,m$ (with $|m|\leq l$), $|l,m\rangle:= U\,[\sqrt{1+1/2}\,Y_l^m(\theta,\phi)]$. If we define
\begin{equation}\label{112}
\widetilde L:= ULU^{-1}\,,\qquad \widetilde M:= UMU^{-1}\,,
\end{equation}
we have\,,
\begin{equation}\label{113}
\widetilde L\,|l,m\rangle=l\,|l,m\rangle\,,\qquad \widetilde M\,|l,m\rangle =m\,|l,m\rangle\,.
\end{equation}
The operators $\widetilde L$ and $\widetilde M$ on $\mathcal H$, as well as $L$ and $M$ on $L^2(S^2,d\Omega)$ are obviously unbounded and self-adjoint on its maximal domain as symmetric generators of a Lie algebra. Next, we are going to construct a RHS on which they are, in addition, continuous \cite{CGO2}. Let us consider the subspace $\Phi$ of all vectors
 $\ds |f\rangle= \sum_{l=0}^\infty \sum_{m=-l}^l f_{l,m}\,|l,m\rangle\in\mathcal H$,  such that
\begin{equation}\label{114}
\big|\big|\,|f\rangle\big|\big|^2_p:=\sum_{l=0}^\infty \sum_{m=-l}^l \big| f_{l,m}\big|^2 (l+|m|+1)^{2p}\,,\qquad p=0,1,2,\dots\,.
\end{equation}
The objects $\big|\big|-\big|\big|_p$ are indeed norms, which provides $\Phi$ of a metrizable locally convex topology. For $p=0$, we have the norm on $\mathcal H$, so that the canonical injection $i:\Phi\longmapsto \mathcal H$ is continuous. Take the anti-dual space $\Phi^\times$ and endow it with the weak topology compatible with the dual pair $\{\Phi,\Phi^\times\}$. Thus, we have the RHS:
\begin{equation}\label{115}
\Phi\subset \mathcal H \subset \Phi^\times\,.
\end{equation}
Then, define $\mathcal D:= U^{-1}\Phi$, and transport the topology  from $\Phi$ to $\mathcal D$. This topology is given by the norms
\begin{equation}\label{116}
\big|\big|f(\theta,\phi)\big|\big|_p^2= \sum_{l=0}^\infty \sum_{m=-l}^l \big| f_{l,m}\big|^2 (l+|m|+1)^{2p}\,,\qquad p=0,1,2,\dots\,.
\end{equation}

The anti-dual $\mathcal D^\times=U^{-1}\Phi^\times$ is defined via the extension of $U^{-1}$ to $\Phi^\times$ via a duality formula of the type \eqref{8}. We have the rigged Hilbert space
\begin{equation}\label{117}
\mathcal D\subset L^2(S^2,d\Omega) \subset\mathcal D^\times\,,
\end{equation}
unitarily equivalent to  $\Phi\subset \mathcal H \subset \Phi^\times\,$ \eqref{115}.
\subsection{Continuous bases depending on the angular variables}

Let us begin with $f(\theta,\phi)\in\mathcal D$ and $|f\rangle =U[f(\theta,\phi)]\in\Phi$. For fixed angles with values $0\le \theta<\pi$, $0\le\phi<2\pi$, almost elsewhere, define the following continuous anti-linear functional, $|\theta,\phi\rangle$, on $\Phi$: For arbitrary $|f\rangle\in\Phi$, one defines the mapping $|\theta,\phi\rangle$ as
\begin{equation}\label{118}
\langle f|\theta,\phi\rangle:= \langle \theta,\phi|f\rangle^* :=f(\theta,\phi)\,,
\end{equation}
where the star denotes complex conjugation. The linearity of each $|\theta,\phi\rangle$ on $\Phi$ is obvious. In order to prove the continuity, take,
\begin{equation}\begin{array}{lll}\label{119}
\ds \langle f|\theta,\phi\rangle =f(\theta,\phi)&=&\ds\sum_{l=0}^\infty \sum_{m=-l}^l f_{l,m}\,\sqrt{l+1/2}\,Y_l^m (\theta,\phi)  \nonumber\\[0.4cm]
& =&\ds \sum_{l=0}^\infty \sum_{m=-l}^l f_{l,m} (l+|m|+1)^p \left( \frac{\sqrt{l+1/2}}{(l+|m|+1)^p}\, Y_l^m (\theta,\phi) \right)\,,
\end{array}\end{equation}
where $p$ is a natural number with $p\ge 3$. Then, take the modulus in \eqref{119} and use the Schwarz inequality in the right hand side. We have
\begin{equation}\label{120}
\big|\langle f\big|\theta,\phi\rangle\big| \le \sqrt{\sum_{l=0}^\infty \sum_{m=-l}^l \big| f_{l,m}\big|^2\, (l+|m|+1)^{2p}} \times \sqrt{\frac{l+1/2}{(l+|m|+1)^{2p}}\,\big|Y_l^m(\theta,\phi)\big|^2}\,.
\end{equation}
The first factor in the right hand side of \eqref{120} is nothing else than $||\,|f\rangle||_p$, while the second factor converges due to the fact that $|Y_l^m(\theta,\phi)|^2\le (2\pi)^{-1}$  for all $\theta,\phi$ \cite{AH}. If we call $C$ this second factor, we finally conclude that
\begin{equation}\label{121}
|\langle f|\theta,\phi\rangle| \le C\,||\,|f\rangle||_p\,,
\end{equation}
which, after \eqref{4}, guarantees the continuity of the functional $|\theta,\phi\rangle$ on $\Phi$, so that $|\theta,\phi\rangle\in\Phi^\times$ for almost all $0\le \theta<\pi$, $0\le\phi<2\pi$. These functionals have some interesting properties: 

\begin{itemize}

\item{For any $f(\theta,\phi)\in\mathcal D$, we can define the operator $\cos\Theta\,f(\theta,\phi) := \cos\theta\,f(\theta,\phi)$. One has that $\cos\theta\,f(\theta,\phi)\in\mathcal D$ and $\cos\Theta$ is continuous on $\mathcal D$. Therefore, we may define $\widehat {\cos\Theta}:= U\,\cos\Theta\,U^{-1}$, which is a symmetric continuous linear operator on $\Phi$ and, hence, can be extended into the anti-dual $\Phi^\times$ by the duality formula \eqref{6}. For almost all $0\le \theta<\pi$, $0\le\phi<2\pi$, we can prove that 
\begin{equation}\label{122}
\widehat {\cos\Theta} \,|\theta,\phi\rangle =\cos\theta \,|\theta,\phi\rangle\,.
\end{equation}
}
\item{Analogously, if we define the operator $e^{i\Phi}$ on $f(\theta,\phi)\in\mathcal D$ as $e^{i\Phi}\,f(\theta,\phi):= e^{i\phi}\,f(\theta,\phi)$ and $\widehat{e^{i\Phi}}:= U\,e^{i\Phi}\,U^{-1}$, we have that 
\begin{equation}\label{123}
\widehat{e^{i\Phi}}\,|\theta,\phi\rangle =e^{i\phi}\,|\theta,\phi\rangle\,.
\end{equation}
}
\item{Let $|g\rangle,|f\rangle\in\Phi$. Their scalar product is
\begin{equation}\label{124}
\langle g|f\rangle =\langle U^{-1}g|U^{-1}f\rangle =\int_{S^2}d\Omega\, g(\theta,\phi)^*\,f(\theta,\phi) = \int_{S^2} d\Omega\, \langle g|\theta,\phi\rangle\langle \theta,\phi|f\rangle\,.
\end{equation}
Then, we may write the following formal identity:
\begin{equation}\label{125}
\mathbb I=\int_{S^2} d\Omega\,|\theta,\phi\rangle\langle \theta,\phi| =\int_0^{2\pi} d\phi \int_0^\pi d(\cos\theta) \,|\theta,\phi\rangle\langle\theta,\phi|  \,.
\end{equation}
We give below the meaning of this $\mathbb I$. 
}

\end{itemize}

Let us take the formal identity $\mathbb I$ as in \eqref{125} and let us apply it to the arbitrary vector $|f\rangle\in\Phi$. It gives
\begin{equation}\label{126}
|f\rangle=\mathbb I\,|f\rangle = \int_{S^2} d\Omega\,|\theta,\phi\rangle\langle \theta,\phi|f\rangle = \int_{S^2} d\Omega\,  f(\theta,\phi) \,|\theta,\phi\rangle\,.
\end{equation}
This gives a span of $|f\rangle$ in terms of the vectors of the form $|\theta,\phi\rangle$. This justifies the name of {\it continuous basis} for the set of vectors $\{|\theta,\phi\rangle\}$, $0\le \theta<\pi$, $0\le\phi<2\pi$. Furthermore, the formal expression \eqref{126} is indeed a continuous anti-linear functional on $\Phi$. If we apply it to an arbitrary vector $|g\rangle\in\Phi$ and take the modulus, it comes 
\begin{equation}\label{127}
\left|\langle g|\int_{S^2} d\Omega\,|\theta,\phi\rangle\langle \theta,\phi|f\rangle\right| \le \int_{S^2} d\Omega\,\big|\langle g\big|\theta,\phi\rangle\big|\cdot  \big|\langle\theta,\phi|f\rangle\big| \le 4\pi\, C^2\,\big|\big|\,|f\rangle\big|\big|_p\cdot\big|\big|\,|g\rangle\big|\big|_p =K\,\big|\big|g\rangle\big|\big|_p\,,
\end{equation}
with $K= 4\pi\, C^2\,\big|\big|\,|f\rangle\big|\big|_p$. Thus, the right hand side in \eqref{126} makes sense as an element of $\Phi^\times$. Consequently, the identity $\mathbb I$ represents the canonical identity from $\Phi$ into $\Phi^\times$. In particular \eqref{126} gives
\begin{equation}\label{128}
|l,m\rangle= \int_{S^2}d\Omega \, \sqrt{l+1/2} \,Y_l^m(\theta,\phi)\,|\theta,\phi\rangle\,.
\end{equation}
Since $\{|l,m\rangle\}$ is a basis for $\mathcal H$, the identity $\mathcal I$ on $\mathcal H$ may be written as
\begin{equation}\label{129}
\mathcal I=\sum_{l=0}^\infty \sum_{m=-l}^l |l,m\rangle\langle l,m|\,.
\end{equation}
Thus, for each $|f\rangle\in\Phi$, we may write
\begin{equation}\label{130}
f(\theta,\phi)^* =\langle f|\theta,\phi\rangle = (\langle f|\mathcal I)|\theta,\phi\rangle = \sum_{l=0}^\infty \sum_{m=-l}^l \langle f|l,m\rangle\langle l,m|\theta,\phi\rangle = \sum_{l=0}^\infty \sum_{m=-l}^l Y_l^m(\theta,\phi)\,\langle f|l,m\rangle \,,
\end{equation}
so that, omitting the arbitrary $|f\rangle\in\Phi$, we have that
\begin{equation}\label{131}
|\theta,\phi\rangle= \sum_{l=0}^\infty \sum_{m=-l}^l Y_l^m(\theta,\phi)\,|l,m\rangle\,,
\end{equation}
which may be looked as the inversion formula for \eqref{126}\,. If we multiply \eqref{126} to the left by $\langle \theta',\phi'|$, operation which is legitimate, we immediately realize that
\begin{equation}\label{132}
\langle\theta',\phi'|\theta,\phi\rangle =\delta(\cos\theta'-\cos\theta')\,\delta(\phi'-\phi)\,,
\end{equation}
which is a textbook formula. 
\subsection{Continuity of the generators of $so(3,2)$}

Along the present section, we are going to use the following definitions for the generators of the $so(3,2)$ Lie algebra, based on the action of these generators on the spherical harmonics \cite{CO}:
\begin{equation}\begin{array}{lll}\label{133}
J_\pm\,Y_l^m(\theta,\phi) &:=&\ds \sqrt{(l\mp m)(l\pm m+1)}\, Y_l^{m+1}(\theta,\phi)\,,\\[0.4cm] 
K_\pm\, Y_l^m(\theta,\phi) &:=&\ds \sqrt{\left(l+\frac12\pm\frac12\right)^2-m^2}\,Y_{l+1}^m(\theta,\phi)\,, \\[0.4cm] 
R_\pm\, Y_l^m(\theta,\phi) &:=&\ds \sqrt{(l+m+1\pm 1)(l+m\pm 1)} \, Y_{l+1}^{m+1}(\theta,\phi)\,,\\[0.4cm] 
S_\pm\, Y_l^m(\theta,\phi) &:=&\ds \sqrt{(l-m+1\pm 1)(l-m\pm 1)}\, Y_{l+1}^{m-1}(\theta,\phi)\,.
\end{array}\end{equation}
These operators can be extended to closed linear operator on suitable dense domains. In addition, we have the generators of the Cartan subalgebra, since the rank of the $so(3,2)$ Lie algebra is 2 and and its dimension is 10. These generators are precisely the operators $L$ and $M$ defined in \eqref{111}. Correspondingly, we have analogous operators densely defined on $\mathcal H$ as
\begin{equation}\label{134}
\widehat J_\pm = U J_\pm U^{-1}\,,\;\; \widehat K_\pm = UK_\pm U^{-1}\,, \;\; \widehat R_\pm = UR_\pm U^{-1}\,,\;\; \widehat S_\pm = US_\pm U^{-1}\,.
\end{equation}
The action of operators \eqref{134} on the elements of the basis $\{|l,m\rangle\}$ is obvious. The continuity of these operators on $\Phi$ and $\Phi^\times$ has been established in \cite{CGO2}. For instance, assume that $|f\rangle=\sum_{l=0}^\infty\sum_{m=-l}^l f_{l,m}\,|l,m\rangle \in\Phi$. Then, write
\begin{eqnarray}\label{135}
\widehat L|f\rangle = \sum_{l=0}^\infty\sum_{m=-l}^l l\, f_{l,m}\,|l,m\rangle\,,
\end{eqnarray}
and
\begin{equation}\begin{array}{lll}\label{136}
\ds \big|\big|\widehat L|f\rangle\big|\big|^2_p &=& \ds \sum_{l=0}^\infty\sum_{m=-l}^l l^2\, \big| f_{l,m}\big|^2 (l+|m|+1)^{2p} 
\\[0.4cm]
 &\le &\ds \sum_{l=0}^\infty\sum_{m=-l}^l  \big| f_{l,m}\big|^2 (l+|m|+1)^{2p+2} =\big|\big|\,|f\rangle\big|\big|^2_{p+1}\,,
\end{array}\end{equation}
expression valid for $p=0,1,2,\dots$. This means that $\widehat L|f\rangle\in\Phi$ if $|f\rangle\in\Phi$. Due to \eqref{5}, \eqref{136} and the linearity of $\widehat L$, it is continuous. Since $\widehat L$ is symmetric, it is extensible to $\Phi^\times$ with continuity under the weak topology.  

The proof for the continuity of the operators in \eqref{134} on $\Phi$ is similar. In order to extend these operators by continuity to $\Phi^\times$, we have to realize first that all the operators with index $+$ are the formal adjoints of the corresponding operator with sign $-$ and viceversa, for instance $\widehat K_+$ and $\widehat K_-$ are formal adjoint of each other. Therefore, to extend these operators to $\Phi^\times$, we only have to use the duality formual \eqref{6}. Needless to say that $L$, $M$ and operators \eqref{133} have the same properties on $\mathcal D \subset \mathcal H\subset \mathcal D^\times$. 


\section{The $su(1,1)$ Lie algebra and Laguerre functions}\label{su11laguerre}

The associated Laguerre polynomials with index $\alpha\in (-1,\infty)$, $L_n^\alpha(y)$, $n=0,1,2,\dots$, are defined on the half-line $\mathbb R^+ \equiv [0,\infty)$  \cite{SZE,AS72,OLBC}. An orthonormal basis on the Hilbert space $L^2(\mathbb R^+)$ is given by the following functions
\begin{equation}\label{137}
M_n^\alpha (y):= \sqrt{\frac{\Gamma(n+1)}{\Gamma(n+a+1)}}\, y^{\alpha/2}\,e^{-y/2}\, L_n^\alpha(y)\,, \qquad n=0,1,2,\dots\,,
\end{equation}
and $\alpha$ fixed. 

Let us consider the space, $\mathcal D_\alpha$, of vectors $f(y)=\sum_{n=0}^\infty a_n\,M_n^\alpha(y)\in L^2(\mathbb R^+)$, such that

\begin{equation}\label{138}
||f(y)||_p^2:= \sum_{n=0}^\infty |a_n|^2 (n+1)^{2p} (n+\alpha+2)^{2p}<\infty, \qquad p=0,1,2,\dots\,,
\end{equation}
with the topology produced by the norms $||-||_p$. With this topology, the space $\mathcal D_\alpha$ is a Fr\'echet nuclear space and is dense in $L^2(\mathbb R^+)$. For $p=0$, we have the Hilbert space norm, so that the canonical injection 
$  i : \mathcal D_\alpha\longmapsto \mathcal H$ is continuous. In consequence, for any fixed $\alpha\in (-1,\infty)$,

\begin{equation}\label{139}
\mathcal D_\alpha\subset L^2(\mathbb R^+) \subset \mathcal D_\alpha^\times\,,
\end{equation}
is a  RHS. 


\subsection{Symmetries of the Laguerre functions}

The following operators defined on the functions $M_n^\alpha(y)$ as
\begin{equation}\label{140}
Y\,M_n^\alpha(y):=y\,M_n^\alpha(y)\,,\quad D_y\,M_n^\alpha(y):= \frac{d}{dy}\, M_n^\alpha(y)\,,\quad N\,M_n^\alpha(y):= n\,M_n^\alpha(y)\,,
\end{equation}
admit closed extensions on $L^2(\mathbb R^+)$. In addition, define the following operators \cite{CO17}:
\begin{equation}\label{141}
K_\pm:= \pm Y\,D_y+N+I+\frac{\alpha-Y}{2}\,, 
\qquad K_3:= N+\frac{\alpha+1}{2}\,I\,,
\end{equation}
where $I$ is the identity operator. The action of these operators on the functions of the basis $\{M_n^\alpha(y)\}$ is
\begin{equation}\begin{array}{l}\label{142}
K_\pm \, M_n^\alpha(y) =\ds \sqrt{(n+\frac12\pm \frac12)(n+\alpha +\frac12\pm \frac12)}\,M_{n+1}^\alpha(y)\,, 
\\[0.4cm] 
K_3\,M_n^\alpha(y) = \ds (n+(\alpha+1)/2)\,M_n^\alpha(y)\,.
\end{array}\end{equation}
Note that $K_+$ and $K_-$ are the formal adjoint of each other (i.e. $(K_\pm)^\dagger =K_\mp$ ) and
\begin{equation}\label{144}
Y = -(K_++K_-) + 2N +(\alpha+1)I\,.
\end{equation}
The commutation relations of $K_\pm$ and $K_3$ are
\begin{equation}\label{145}
[K_3,K_\pm]=\pm K_\pm\,,\qquad [K_+,K_-]=-2K_3\,,
\end{equation}
which are the commutation relations for the generators of the $su(1,1)$ Lie algebra \cite{CO13}. The Casimir is
\begin{equation}\label{146}
\mathcal C=K_3^2-\frac 12 \,\{K_+,K_-\}=\frac{\alpha^2-1}{4}\,I\,.
\end{equation}
The next result concerns the continuity of these operators. 

{\bf Proposition 2.-}{\it  

The operators $K_\pm$, $K_3$, $Y$ and $Y\,D_y$ are continuous on $\mathcal D_\alpha$ for fixed $\alpha\in(-1,\infty)$. 
}\smallskip

{\bf Proof.-} 
Let $f(y)=\sum_{n=0}^\infty a_n\,M_n^\alpha(y)\in\mathcal D_\alpha$. Then,
\begin{equation}\label{147}
K_+\,f(y)= \sum_{n=0}^\infty \sqrt{(n+1)(n+\alpha+1) }\, a_n\,M_n^\alpha(y)\,.
\end{equation}
We need to show that \eqref{147} is well defined on $\mathcal D_\alpha$. For all $p=0,1,2,\dots$, take,
\begin{equation}\begin{array}{lll}\label{148}
\big|\big| K_+\,f(y)\big|\big|_p^2 &=&\ds \sum_{n=0}^\infty (n+1)(n+\alpha+1)(n+1)^{2p}(n+\alpha+2)^{2p}\,\big|a_n\big|^2 
\\[0.4cm] 
&\le&\ds \sum_{n=0}^\infty (n+1)^{2p+2}(n+\alpha+1)^{2p+2} \,\big|a_n\big|^2 \le \big|\big|f(y)\big|\big|_{p+1}^2\,.
\end{array}\end{equation}
This shows both our claim and the continuity of $K_+$ on $\mathcal D_\alpha$. Proofs for $K_-$ and $K_3$ are similar. The continuity of $Y$ comes from \eqref{144} and the continuity of $Y\,D_y$ from \eqref{141}.
\hfill$\blacksquare$ 


\subsection{RHS and continuous bases}

In order to define the continuous basis, we need an abstract RHS, which is the usual procedure. Let us consider an abstract infinite dimensional separable Hilbert space $\mathcal H$ and a unitary operator $U: \mathcal H=L^2(\mathbb R^+)$. We choose the operator $U$ as that given by the Gelfand-Maurin theorem (section \ref{riggedhs}), where the role of $A$ is played by  the operator $Y$ defined in \eqref{144}. This unitary operator $U$ is not necessarily unique, although this is irrelevant, choose any one that makes this job. 

Then, define for each $\alpha\in (-1,\infty)$ the space $\Phi_\alpha:= U^{-1}\mathcal D_\alpha$ and transport the topology from $\mathcal D_\alpha$ to $\Phi_\alpha$ by $U^{-1}$. Call $|n,\alpha\rangle:= U^{-1}\,M_n^\alpha(y)$. For any $|f\rangle =\sum_{n=0}^\infty a_n\,|n,\alpha\rangle\in\Phi_\alpha$, the norms defining the topology are
\begin{equation}\label{149}
\big|\big|\,|f\rangle\big|\big|_p^2:= \sum_{n=0}^\infty \big|a_n\big|^2\, (n+1)^{2p}(n+\alpha+1)^{2p}\,,\qquad p=0,1,2,\dots\,.
\end{equation}

We have the family of RHS given by $\Phi_\alpha\subset \mathcal H\subset \Phi_\alpha^\times$ for each $\alpha\in(-1,\infty)$. Let us define the operator $\widetilde Y:= UYU^{-1}$, which is continuous on each of the $\Phi_\alpha$.  After the Gelfand-Maurin theorem, we conclude that there exists a set of functionals $|y\rangle\in\Phi_\alpha^\times$ for $y\in\mathbb R^+$, such that $\widetilde Y\,|y\rangle =y\,|y\rangle$. In the kets $|y\rangle$, we omit the index $\alpha$ for simplicity. Furthermore and according to \eqref{6}, for any pair of vectors $|f\rangle,|g\rangle\in\Phi_\alpha$, we have that
\begin{equation}\label{150}
\langle f|g\rangle =\int_0^\infty \langle f|y\rangle\langle y|g\rangle\,dy\,, \qquad \langle f|n,\alpha\rangle =\int_0^\infty \langle f|y\rangle\langle y|n,\alpha\rangle\,dy = \int_0^\infty f^*(y)\,M_n^\alpha(y)\,dy\,.
\end{equation}
If we omit the arbitrary bra $\langle f|$ in both identities of \eqref{150}, we obtain the following information:
\begin{enumerate}

\item 
 For each $|g\rangle\in\Phi_\alpha$, we have the decomposition
\begin{equation}\label{151}
|g\rangle =\int_0^\infty |y\rangle\langle y|g\rangle\,dy =\int_0^\infty g(y)\,|y\rangle\,dy\,,
\end{equation}
which shows that the functionals $|y\rangle$, for all $y\in\mathbb R^+$, form a continuous basis for $\Phi_\alpha$.

\item 
 Vectors in the discrete and continuous basis are related by
\begin{equation}\label{152}
|n,\alpha\rangle =\int_0^\infty M_n^\alpha(y)\,|y\rangle\,dy\,.
\end{equation}

\item 
 If, in addition, we omit the arbitrary ket $|g\rangle$ in \eqref{150}, we obtain the following identity,
\begin{equation}\label{153}
\mathbb I =\int_0^\infty |y\rangle\langle y|\,dy\,,
\end{equation}
which is the canonical injection $\mathbb I: \Phi_\alpha\longmapsto \Phi_\alpha^\times$. 
\end{enumerate}
\section{The $su(2,2)$ Lie algebra and algebraic Jacobi functions}\label{algebraicjacobifunctions}

The Jacobi polynomials of order $n\in\mathbb N$, $J^{\alpha,\beta}_n(x)$, are usually defined as

\begin{equation}\label{154}
J^{(\alpha,\beta)}_n(x) :=  \sum_{s=0}^n \left(\begin{array}{c} n+\alpha \\ s \end{array}\right) \left(\begin{array}{c} n+\beta \\ n-s \end{array}\right) \left(\frac{x+1}{2} \right)^s \left(\frac{x-1}{2} \right)^{n-s}\,,
\end{equation}
with
\begin{equation}\label{155}
\left(\begin{array}{c} a \\ s \end{array} \right):= \frac{(a+1-s)(a+1-s+1)\dots a}{s!}\,,
\end{equation}
which are the generalized binomial coefficients, $a$ is an arbitrary number and $s$ a positive integer  \cite{SZE,AS72,OLBC}. They verify the following second order differential equation:
\begin{equation}\label{156}
\left[ (1-x^2) \frac{d^2}{dx^2} -((\alpha+\beta+2)x+(\alpha-\beta)) \frac{d}{dx} + n(n+\alpha+\beta+1)\right]\,
J_n^{(\alpha,\beta)}(x)=0 \,.
\end{equation}

\subsection{Algebraic Jacobi functions}

Jacobi polynomials yield to the main concept of this section, the {\it algebraic Jacobi functions} \cite{CO,CGO3}, defined as
\begin{equation}\label{157}
\mathcal J^{m,q}_j(x):= \sqrt{\frac{\Gamma(j+m+1) \Gamma(j-m+1)}{\Gamma(j+q+1) \Gamma(j-q+1)}} \left(\frac{1-x}{2} \right)^{\frac{m+q}{2}} \left(\frac{1+x}{2} \right)^{\frac{m-q}{2}} J^{(m+q,m-q)}_{j-m}(x)\,,
\end{equation}
where,
\begin{equation}\label{158}
j:= n+\frac{\alpha+\beta}{2}\,, \qquad m:= \frac{\alpha+\beta}{2}\,,\qquad \beta:= m-q\,.
\end{equation}
Considerations derived from the theory of group representations force the following restrictions  in the above parameters:
\begin{equation}\label{159}
j \ge |m|\,, \quad j\ge |q|\,, \quad 2j\in\mathbb N\,, \quad  j-m\in\mathbb N\,, \quad  j-q\in\mathbb N\,,
\end{equation}
and the parameters $(j,m,q)$ are all together integers or half-integers. We may rewrite conditions \eqref{159} in terms of the original parameters $(n,\alpha,\beta)$ as
\begin{equation}\label{160}
n\in\mathbb N\,,\quad \alpha,\beta \in \mathbb Z\,, \quad \alpha\ge -n\,, \quad \beta \ge -n\,, \quad \alpha+\beta \ge-n\,.
\end{equation}
The  algebraic Jacobi functions ${\cal J}_j^{m,q}(x)$ verify the following differential equation:
\begin{equation}\label{161}
\left[-(1-x^2)\,\frac{d^2}{dx^2}+2\,x\frac{d}{dx}+ 
\frac{2\;m\;q\;x+m^2+q^2}{1-x^2}- j(j+1)\right] \,{\cal J}_j^{m,q}(x)=0\,,
\end{equation} 
where the symmetry under the interchange $m \Leftrightarrow q$ is evident. 
In addition, for fixed $m$ and $q$ the algebraic Jacobi functions satisfy the following relations:
\begin{equation}\begin{array}{rll}\label{162}
\ds \int_{-1}^1 \mathcal J_j^{m,q}(x) (j+1/2) \, \mathcal J_{j'}^{m,q}(x)\,dx &=&\delta_{jj'}\,,\\[0.4cm]
\ds \sum_{j\ge \sup(|m|,|q|)}^\infty \mathcal J_j^{m,q}(x) (j+1/2) \, \mathcal J_{j}^{m,q}(y) &=& \delta(x-y)\,.
\end{array}\end{equation}
The indices $j$, $m$ and $q$ are either integer or half-integer. Relations \eqref{162}  show that for $m$ and $q$ being fixed, the set of functions given by $\{\sqrt{j+1/2}\,J_j^{m,q}(x)\}_{j\ge \sup(|m|,|q|)}^\infty$ forms an orthonormal basis of the Hilbert space $L^2[-1,1]$.

We may comment in passing the existence of a relation  between the Legendre functions and some of the algebraic Jacobi functions, which is
\begin{equation}\label{164}
P_l^m(x) =(-1)^m \sqrt{\frac{(l+m)!}{(l-m)!}}\,\mathcal J_l^{m,0}(x)\,.
\end{equation}

\subsection{Symmetries of the algebraic Jacobi functions}

Also, the ladder operators, $A_\pm,B_\pm,C_\pm,D_\pm,E_\pm,F_\pm$  that appear in the theory of algebraic Jacobi functions are generators of the $su(2,2)$ Lie algebra \cite{CO2,COV19}. Their action on the algebraic Jacobi functions is given by
\begin{equation}\begin{array}{lll}\label{165}
A_\pm\, \mathcal J^{m,q}_j(x)&=& \sqrt{(j\mp m)(j\pm m+1)}\,\mathcal J^{m\pm 1,q}_j(x)\,,
\\[0.4cm]
B_\pm \, \mathcal J^{m,q}_j(x)&=& \sqrt{(j\mp q)(j\pm q+1)}\, \mathcal J^{m,q\pm 1}_j(x)\,,
\\[0.4cm]
C_\pm\,{\cal J}_j^{m,q}(x)&=&\sqrt{(j+m+\frac12\pm\frac12)(j+q+\frac12\pm\frac12)}\; {\cal J}_{j\pm 1/2}^{m\pm 1/2,\;q\pm 1/2}(x),\\[0.4cm]
D_\pm\,{\cal J}_j^{m,q}(x)&=&\sqrt{(j+m+\frac12\pm\frac12)(j-q+\frac12\pm\frac12)}\; {\cal J}_{j\pm1/2}^{m\pm1/2,\;q\mp 1/2}(x)\,\\[0.4cm]
E_\pm\,{\cal J}_j^{m,q}(x)&=&\sqrt{(j-m+\frac12\pm\frac12)\,(j+q+\frac12\pm\frac12)}
\;{\cal J}_{j\pm 1/2}^{m\mp 1/2,\;q\pm 1/2}(x),\\[0.4cm]
F_\pm\,{\cal J}_j^{m,q}(x)&=&\sqrt{(j-m+\frac12\pm\frac12)\,(j-q+\frac12\pm\frac12)}\;
\,{\cal J}_{j\pm 1/2}^{m\mp 1/2,\,q\mp 1/2}(x)\,.
\end{array}\end{equation}
The generators of the Cartan subalgebra, $J$, $M$ and $Q$, act on the algebraic Jacobi functions as follows:
\begin{equation}\label{166}
J\,{\cal J}_j^{m,q}(x) =j\,{\cal J}_j^{m,q}(x)\,,\quad M\,{\cal J}_j^{m,q}(x)=m\,{\cal J}_j^{m,q}(x)\,,\quad Q\,{\cal J}_j^{m,q}(x)=q\, {\cal J}_j^{m,q}(x)\,.
\end{equation}
All these operators can be extended to unbounded closed operators on $L^2[-1,1]$. In the case of $J$, $M$ and $Q$, they admit self-adjoint extensions. Operators denoted with the same capital letter and different sign are formal adjoint (conjugate Hermitian) of each other (i.e., $(X_\pm )^\dagger=X_\mp$). On functions ${\cal J}_j^{m,q}(x)$ with $j\geq|m| > |q|$, one may define the following  pair of mutually Hermitian formal adjoint operators:
\begin{equation}\label{167}
K_\pm:=   F_\pm C_\pm  \frac{1}{\sqrt{(J+1/2\pm 1/2)^2-Q^2}}\,, 
\end{equation}
so that
\begin{equation}\label{168}
 K_\pm\;  {\cal J}_j^{m,q}(x) = \sqrt{(j+1/2\pm 1/2)^2-m^2}\,  {\cal J}_{j\pm1}^{m,q}(x)\,.
\end{equation}
These operators $K_\pm$ along $K_3:= J+1/2$ close a $su(1,1)$ Lie algebra, since:
\begin{equation}\label{169}
[K_+,K_-]=-2 K_3,\qquad [K_3,K_\pm]=\pm K_\pm\,,
\end{equation}
and the set of functions $\{{\cal J}_j^{m,q}(x)\}^{m,q\, {\rm fixed}}_{j\geq |m| > |q|}$ with $|m|>|q|$ is a basis of the space supporting a UIR of the group $SU(1,1)$ with Casimir $\mathcal C=m^2-1/4$. 

\subsection{Algebraic Jacobi functions on the hypersphere $\mathcal S^3$}\label{hyperesfere}

In the precedent analysis, we have deal with situations in which the number of discrete and continuous variables is the same. This idea revealed to be of importance in the analysis of the spaces which make continuous the above operators, if we are really interested in a description encompassing the maximal generality. To this end, we define the following functions:
\begin{equation}\label{170}
\mathcal N^{m,q}_j(x,\phi,\chi) := \sqrt{j+1/2}\,\mathcal J^{m,q}_j(x)\,e^{im\phi}\,e^{iq\chi}\,, 
\end{equation}
where $\phi$ and $\chi$ are two angular variables, $\phi\in[0,2\pi)$ and $\chi\in[0, \pi]$ ($x$ could be considered as $x=\cos \theta$ with $\theta \in [0,\pi]$ and in  this case the Jacobi functions will live in the hypersphere $\mathcal S^3$). Thus, the $\mathcal N$-functions defined in \eqref{170} depend on the variables, $x,\phi,\chi$, and the discrete parameters $j,m,q$. The properties of the Jacobi functions yield to the following orthogonality relations valid for for {\it either} $j$ integer {\it or} $j$ half-integer, with $m,q=-j,-j+1,\dots,j-1,j$ and $m',q'= -j',-j'+1,\dots,j'-1,j'$:
\begin{equation}\label{171}
\frac 1{2\pi^2} \int_0^{2\pi} d\phi \int_0^{\pi} d\chi \int_{-1}^1 dx \;\mathcal N^{m,q}_j(x,\phi,\chi)\,{\mathcal N^{m',q'}_{j'}}^*(x,\phi,\chi) =\delta_{jj'}\,\delta_{mm'}\,\delta_{qq'}\,.
\end{equation}
These functions satisfy a completeness relation of the type:
\begin{equation}\label{172}
\sum_{j_{min}}^\infty \sum_{m,q=-j}^j |\mathcal N^{m,q}_j(x,\phi,\chi)|^2= I\,,
\end{equation}
where $j_{min}=0$ for integers values of $j$ or  $j_{min}=1/2$ for half-integers and $I$ is an identity. Note that we have two different situations, one when $j$ is an integer and the other when $j$ is a half-integer. In both cases, either $e^{im\phi}$ or $e^{iq\chi}$ span respective vector spaces of dimension $2j+1$. This spaces, being isomorphic to $\mathbb C^{2j+1}$, may be identified with it. Then, for either $j$ integer or half-integer, the set of functions $\mathcal N^{m,q}_j(x,\phi,\chi)$ with $m,q=-j,-j+1,\dots,j-1,j$ is the basis for the following  Hilbert spaces:

\begin{equation}\label{173}
\mathcal H_I:= \bigoplus_{j=0}^\infty L^2[-1,1]\otimes \mathbb C^{2j+1}\otimes \mathbb C^{2j+1}\,,\qquad \mathcal H_H:=\bigoplus_{j=1/2}^\infty L^2[-1,1]\otimes \mathbb C^{2j+1}\otimes \mathbb C^{2j+1}\,,
\end{equation}
respectively. The subindices $I$ and $H$ stand for integer and half-integer, respectively. Then, let us take $f_I(x,\phi,\chi)\in\mathcal H_I$ and $f_H(x,\phi,\chi)\in\mathcal H_H$, so that
\begin{equation}\label{174}
f_I(x,\phi,\chi) = \sum_{j=0}^\infty \sum_{m,q=-j}^j a_{j,m,q}\,  \mathcal N_j^{m,q}(x,\phi,\chi)\,, \quad f_H(x,\phi,\chi) = \sum_{j=1/2}^\infty \sum_{m,q=-j}^j b_{j,m,q}\,  \mathcal N_j^{m,q}(x,\phi,\chi)\,.
\end{equation}

\subsection{RHS associated to the algebraic Jacobi functions}

Next, we define two new rigged Hilbert spaces. The spaces of test functions $\Phi_I$ and $\Phi_H$ are the functions in $\mathcal H_I$ and $\mathcal H_H$, respectively, such that
\begin{equation}\label{175}
\left[p^{I}_{r,s}(f_I)\right]^2:=\sum_{j=0}^\infty \sum_{m,q=-j}^j |a_{j,m,q}|^2\,(j+|m|+1)^{2r}\,(j+|q|+1)^{2s} <\infty\,,
\end{equation}
and
\begin{equation}\label{176}
\left[p^{H}_{r,s}(f_H)\right]^2:=\sum_{j=1/2}^\infty \sum_{m,q=-j}^j  |b_{j,m,q}|^2\, (j+|m|+1)^{2r}\,(j+|q|+1)^{2s} <\infty\,,
\end{equation}
respectively, with $r,s=0,1,2,\dots$. Observe that both \eqref{175} and \eqref{176} define norms on $\mathcal H_I$ and $\mathcal H_H$, respectively, and they generate respective topologies on $\Phi_I$ and $\Phi_H$. For $r=s=0$, we recover the Hilbert space topology, which shows that the canonical injections $\Phi_{I,H}\longmapsto \mathcal H_{I,H}$ are continuous, so that
\begin{equation}\label{177}
\Phi_I\subset \mathcal H_I \subset \Phi_I^\times\,, \qquad {\rm and}\qquad \Phi_H\subset \mathcal H_H \subset \Phi_H^\times\,,
\end{equation}
are rigged Hilbert spaces. 

Analogously, we define the spaces $\Xi_I$ and $\Xi_H$ as the spaces of functions in $\mathcal H_I$ and $\mathcal H_H$ verifying the following relations:
\begin{equation}\label{178}
t^I_{r,s}(f_I) := \sum_{j=0}^\infty \sum_{m,q=-j}^j |a_{j,m,q}|\,(j+|m|+1)^{r}\,(j+|q|+1)^{s} <\infty\,,
\end{equation}
and
\begin{equation}\label{179}
t^H_{r,s}(f_H) := \sum_{j=1/2}^\infty \sum_{m,q=-j}^j |b_{j,m,q}|\,(j+|m|+1)^{r}\,(j+|q|+1)^{s} <\infty\,,
\end{equation}
respectively, with $r,s=0,1,2,\dots$. These are also norms that endow respective topologies on $\Xi_I$ and $\Xi_H$. 
Since,
\begin{equation}\label{180}
p_{r,s}^{I,H}(f_{I,H}) \le t^{I,H}_{r,s}(f_{I,H}) \,, \qquad r,s=0,1,2,\dots\,,
\end{equation}
we conclude that $\Xi_{I,H}\subset \Phi_{I,H}$ and that the canonical injections $\Xi_{I,H}\longmapsto \Phi_{I,H}$ are continuous. Thus,  we have two new RHS's, and, in addition, we have the following subordinate relations with continuity
\begin{equation}\label{181}
\Xi_{I,H} \subset \Phi_{I,H} \subset \mathcal H_{I,H} \subset \Phi_{I,H}^\times \subset \Xi_{I,H}^\times\,,
\end{equation}
where in each sequence in \eqref{181}, we should keep either the subindex $I$ or $H$. 
\subsection{Continuity of the $su(2,2)$ operators}

The operators $J$, $M$ and $Q$, defined above in this section, admit obvious extensions to respective dense subspaces of $\mathcal H_{I,H}$. For instance,
\begin{equation}\label{182}
(J f_I)(x,\phi,\chi) = \sum_{j=0}^\infty \sum_{m,q=-j}^j j\,a_{j,m,q}\,  \mathcal N_j^{m,q}(x,\phi,\chi)\,.
\end{equation}
Thus,
\begin{equation}\begin{array}{lll}\label{183}
\left[p_{r,s}^I(Jf_I)\right]^2 &=&\ds \sum_{j=0}^\infty \sum_{m,q=-j}^j  |a_{j,m,q}|^2 \,j^2 \,(j+|m|+1)^{2r}\,(j+|q|+1)^{2s}  
\\[0.4cm] 
&\le&\ds  \sum_{j=0}^\infty \sum_{m,q=-j}^j  |a_{j,m,q}|^2 \, (j+|m|+1)^{2(r+1)}\,(j+|q|+1)^{2s} =[p_{r+1,s}^I(f_I)]^2\,,
\end{array}\end{equation}
for $r,s=0,1,2,\dots$, which proves that $J\Phi_I\subset \Phi_I$ with continuity. Analogously,
\begin{equation}\begin{array}{lll}\label{184}
t_{r,s}^I(Jf_I) &=&\ds \sum_{j=0}^\infty \sum_{m,q=-j}^j  |a_{j,m,q}| \,j \,(j+|m|+1)^{r}\,(j+|q|+1)^{s}  \\[0.4cm] 
&\le&\ds \sum_{j=0}^\infty \sum_{m,q=-j}^j  |a_{j,m,q}| \, (j+|m|+1)^{r+1}\,(j+|q|+1)^{s} = t^I_{r+1,s}(f_I)\,, 
\end{array}\end{equation}
for $r,s=0,1,2,\dots$, which proves that $J\Xi_I\subset \Xi_I$ with continuity. Same for $J$ on $\Phi_H$ and $\Xi_H$ and for $M$ and $Q$ in these four spaces. Since these operators are symmetric and self-adjoint on a proper domain, they may be extended by continuity to the duals. A similar proof is also valid to show the continuity of the ladder operators $A_\pm$ and $B_\pm$, defined in \eqref{165} and $K_\pm$ in \eqref{167} on all the spaces $\Xi_{I,H}$ and $\Phi_{I,H}$ and therefore their extensions by continuity to the duals. 

However, the ladder operators $C_\pm,D_\pm,E_\pm,F_\pm$ have a different nature, as they transform algebraic Jacobi functions of integer indices into the same type of functions with half-integer indices and viceversa.   Under the assumption that $C_-\,\mathcal N_j^{0,0}(x,\phi,\chi)=0$ and the same for $D_-$, $E_-$ and $F_-$, we can easily prove that all these operators are continuous from $\Phi_I$ into $\Phi_H$ and viceversa and the same from $\Xi_I$ into $\Xi_H$ and viceversa. As they are the formal adjoint of each other, we conclude that they can be also continuously extended as analogous relations between the duals. 
\subsection{Discrete and continuous basis}

In the sequel, we omit the subindices $I$ and $H$ for simplicity. All results will be valid for both cases. 
As we have done in all precedent examples, let us consider an abstract infinite dimensional separable Hilbert space $\mathcal G$ and a unitary mapping $U:\mathcal G\longmapsto \mathcal H$. As a matter of fact, there are two of each: $U_{I,H}:\mathcal G_{I,H}\longmapsto \mathcal H_{I,H}$, although we omit the subindices, as we said. Take $\Theta:=U^{-1}\Xi$ and $\Psi:=U^{-1}\Phi$, and endow $\Theta$ and $\Phi$ with the topologies transported by $U^{-1}$ from $\Xi$ and $\Phi$, respectively. Then, we have two new RHS's, $\Theta\subset\mathcal G \subset \Theta^\times$ and $\Psi\subset\mathcal G \subset  \Psi^\times$. We focus our attention in the former. 

For any $|f\rangle\in\Theta$, we define the action of the ket $|x,m,q\rangle$, $x\in[-1,1]$, $m$ and $q$ being fixed, as
\begin{equation}\label{185}
\langle f|x,m,q\rangle:= \sum_{j_{min}}^\infty  a_{j,m,q}\, \mathcal N_j^{m,q}(x,0,0) = \sum_{j_{min}}^\infty a_{j,m,q}\, \sqrt{j+1/2}\, \mathcal J_j^{m,q}(x)\,.
\end{equation}
This definition shows that $|x,m,q\rangle$ is an anti-linear mapping on $\Theta$, which is also continuous since,
\begin{equation}\begin{array}{lll}\label{186}
\big|\langle f|x,m,q\rangle\big| &\le &\ds \sum_j |a_{j,m,q}| \, (j+|m|+1)^2(j+|q|+1) \\[0.4cm]
& \le &\ds \sum_j  \sum_{m,q=-j}^j |a_{j,m,q}| \, (j+|m|+1)^2(j+|q|+1) =t_{2,1}(|f\rangle)\,,
\end{array}\end{equation}
with 
\begin{equation}\label{187}
U\,|f\rangle = f(x,\phi,\chi)=\sum_{j_{min}}^\infty \sum_{m,q=-j}^j |a_{j,m,q}|\,\mathcal N_j^{m,q}(x,\phi,\chi)\,.
\end{equation}
Next, let us define the kets $|j,m,q\rangle$ for any $j$ and any $m,q=-j,-j+1,\dots,j-1,j$ as 
\begin{equation}\label{188}
|j,m,q\rangle:= U^{-1} \,\mathcal N^{m,q}_j(x,\phi,\chi)\,,
\end{equation}
so that \eqref{185} gives
\begin{equation}\label{189}
\langle j,m',q'|x,m,q\rangle = \sqrt{j+1/2}\,\mathcal  J_j^{m',q'}(x)\,\delta_{mm'}\,\delta_{qq'} =\langle x,m,q|j,m',q'\rangle \,,
\end{equation}
since \eqref{189} is real. Observe that there exists the following formal relation between $|x,m,q\rangle$ and $|j,m,q\rangle$:
\begin{equation}\label{190}
|x,m,q\rangle= \sum_{j_{min}}^\infty |j,m,q\rangle\,\sqrt{j+1/2}\,\mathcal J_j^{m,q}(x)\,.
\end{equation}
This is easily justified by multiplying \eqref{190} by $\langle j,m',q'|$:
\begin{equation*}\begin{array}{lll}\label{191}
\langle j',m',q'|x,m,q\rangle &=&\ds \sum_{j_{min}}^\infty \langle j',m',q'|j,m,q\rangle \sqrt{j+1/2}\,\mathcal J_j^{m,q}(x) \\[0.4cm]
& =&\ds \sum_{j_{min}}^\infty \delta_{j,j'}\, \sqrt{j+1/2}\,\mathcal J_j^{m,q}(x) \,\delta_{mm'}\,\delta_{qq'} = \sqrt{j+1/2}\,\mathcal J_j^{m,q}(x) \,\delta_{mm'}\,\delta_{qq'} \,,
\end{array}\end{equation*}
which coincides with \eqref{189}\,.  There are some other formal relations that can be easily obtain. Proofs are published elsewhere \cite{CGO3}, there are simple notwithstanding. First of all, we have
\begin{equation}\label{192}
\langle x',m',q'|x,m,q\rangle = \sum_j \mathcal N^{m',q}_j (x,\phi,\chi) \, \mathcal N^{m,q}_j(x,\phi,\chi) \, \delta_{mm'}\,\delta_{qq'} = \delta(x-x')\,\,\delta_{mm'}\,\delta_{qq'}\,.
\end{equation}
For any $|f\rangle \in \Theta$,  we have the following relation:
\begin{equation}\label{193}
\langle j',m',q'|f\rangle = \sum_{m,q=-\infty}^\infty \int_{-1}^1 \langle j',m',q'|x,m,q\rangle \,f^{m,q}(x)\,dx\,,
\end{equation}
where if $|f\rangle=\sum_{j=0}^\infty \sum_{m,q=-j}^j a_{j,m,q} \sqrt{j+1/2} \,\mathcal J_j^{m,q}(x)$, we have that
\begin{equation}\label{194}
f^{m,q}(x)=\sum_{j=0}^\infty a_{j,m,q} \sqrt{j+1/2} \,\mathcal J_j^{m,q}(x)\,,
\end{equation}
so that 
\begin{equation}\label{195}
|f\rangle = \sum_{m,q=-\infty}^\infty \int_{-1}^1 |x,m,q\rangle \,f^{m,q}(x)\,dx\,,
\end{equation}
which shows that any $|f\rangle\in\Theta$ may be written formally in terms of the elements of the set of functionals $\{|x,m,q\rangle\}$,  which acquires the category of {\it continuous basis} due to this fact. Here, $x\in[-1,1]$, $m,q$ being the set either of the integers or the half-integers, either positive or negative. 

For $|j,m,q\rangle$, the functions $f^{m,q}(x)$ are equal to $\sqrt{j+1/2}\,\mathcal J_j^{m,q}(x)$, which after \eqref{195} gives
\begin{equation}\label{196}
|j,m,q\rangle =  \sum_{m,q=-\infty}^\infty \int_{-1}^1 |x,m,q\rangle \,\sqrt{j+1/2}\,\mathcal J_j^{m,q}(x)\,dx\,,
\end{equation}
which gives the inversion formula for \eqref{190}. We have completed the relation between discrete and continuous basis. Moreover, note that
\begin{equation}\label{197}
f^{m,q}(x)=\langle x,m,q|f\rangle \,,
\end{equation}
and
\begin{equation}\label{198}
\sum_{m,q=-j}^j \int_{-1}^1 dx\,|x,m,q\rangle\langle x,m,q| = \mathcal I\,,
\end{equation}
where $\mathcal I:\Theta\longmapsto\Theta^\times$ is the canonical injection relating this dual pair.  We close here the discussion on Jacobi algebraic functions.

\section{su(1,1)$\oplus$su(1,1), Zernike functions and RHS}\label{zernikeRHS}

The so called Zernike polynomials were introduced by Zernike in 1934  in connection with some applications in the analysis of optical images \cite{ZER}. These Zernike polynomials $R_n^m(r)$, also called Zernike radial polynomials \cite{BWB}, as usually one takes $0\le r\le 1$ in applications, are the solutions of the differential equation,
\begin{equation}\label{199}
\left[ (1-r^2)\,\frac{d^2}{dr^2} -\left(3r-\frac1r   \right) \frac{d}{dr}+n(n+2)-\frac{m^2}{r^2}  \right] R^m_n(r)=0\,,
\end{equation}
verifying
\begin{equation}\label{200}
R_n^m(1)=1\,, \qquad R_n^m(r)=R_n^{-m}(r)\,.
\end{equation}
Explicitly,
\begin{equation}\label{201}
R_n^m(r)=\sum_{k=0}^{\frac{n-m}{2}} (-1)^k\,\left( \begin{array}{c} n-k\\ k\end{array}\right)\,
\left( \begin{array}{c} n-2k\\ \frac{n-m}{2}- k\end{array}\right)\,r^{n-2k}\,.
\end{equation}
For each value of $m$, Zernike polynomials show orthogonality properties:
\begin{equation}\label{202}
\int_0^1 R^m_n(r)\,R^m_{n'}(r)\,r\, dr= \frac{\delta_{nn'}}{2(n+1)}\,,
\end{equation}
as well as a completeness relation such as
\begin{equation}\label{203}
\sum_{\substack{n=|m| \\ n\equiv m\, {\rm (mod\, 2)}}
} ^\infty  R^m_n(r)\,R^m_n(r')\,(n+1)=\frac{\delta(r-r')}{2r}\,.
\end{equation}
They are also related to the Jacobi polynomials according to the following formula:
\begin{equation}\label{204}
R^m_n(r)= (-1)^{(n-m)/2}\, r^m\, J_n^{(m,0)}(1-2r^2)\,.
\end{equation}
Along Zernike polynomials, there exist the Zernike functions $\mathcal Z_n^m(r,\phi)$, which are defined on the closed unit circle
\begin{equation}\label{205}
\mathcal D\equiv \{(r,\phi)\,,\;\; 0\le r\le 1\,,\;\; \phi\in[0,2\pi)\;\}\,,
\end{equation}
as follows:
\begin{equation}\label{206}
\mathcal Z_n^m(r,\phi) := R_n^m(r)\, e^{im\phi}\,,\qquad n\in\mathbb N\,, \; m\in\mathbb Z\,,
\end{equation}
with the conditions $|m|\le n$ and $\frac{n-|m|}{2}\in\mathbb N$\,.

\subsection{$W$-Zernike functions}

From \eqref{206}, we define the $W$-Zernike functions, $W_{u,v}(r,\phi)$, using the following procedure \cite{CGO4}. First of all, introduce the parameters $u$ and $v$, defined as
\begin{equation}\label{207}
u:= \frac{n+m}{2}\,,\qquad v:= \frac{n-m}{2}\,,
\end{equation}
which are positive integers and independent of each other, $u,v=0,1,2,\dots$. With this notation,
\begin{equation}\label{208}
R_n^m(r) \equiv R_{u+v}^{|u-v|}(r)\,.
\end{equation}
The $W$-Zernike functions, $W_{u,v}(r,\phi)$ are functions on the closed unit circle $\mathcal D$, verifying the relation
\begin{equation}\label{209}
W_{u,v}(r,\phi) = \sqrt{\frac{u+v+1}{\pi}}\, \mathcal Z^{u-v}_{u+v}(r,\phi) = \sqrt{\frac{u+v+1}{\pi}}\, R^{|u-v|}_{u+v}(r)\,e^{i(u-v)\phi}\,.
\end{equation}
In addition the $W$-Zernike functions have some interesting properties:

\begin{itemize}

\item{They are square integrable on $\mathcal D$, so that they belong to the Hilbert space $L^2(\mathcal D,rdr d\phi)\equiv L^2(\mathcal D)$.}\medskip

\item{They fulfil some symmetry relations such as
\begin{equation}\label{210}
W_{v,u}(r,\phi)=W_{u,v}(r,\phi)^*= W_{u,v}(r,-\phi)\,,
\end{equation}
where the star denotes complex conjugation.}
\medskip

\item{They are orthonormal on $L^2(\mathcal D)$:
\begin{equation}\label{211}
\langle W_{u',v'},W_{u,v} \rangle = \int_0^{2\pi} d\phi \int_0^1 dr\,r\, W_{u',v'}(r,\phi)^*\,W_{u,v}(r,\phi) = \delta_{uu'}\,\delta_{vv'}\,,
\end{equation}
where $\langle \cdot,\cdot\rangle$ denotes scalar product on $L^2(\mathcal D)$.
}\medskip

\item{A completeness relation holds:
\begin{equation}\label{212}
\sum_{u,v=0}^\infty W_{u,v}(r,\phi)\,W^*_{u,v}(r',\phi')= \frac{1}{2r}\,\delta(r-r')\,\delta(\phi-\phi')\,.
\end{equation}}

\item{The fact that Zernike polynomials are bounded, $|R_n^m(r)|\le 1$ on the interval $0\le r\le 1$, implies an interesting upper bound for the $W$-Zernike functions:
\begin{equation}\label{213}
\big|W_{u,v}(r,\phi)\big|\le \sqrt{\frac{u+v+1}{\pi}}\,,\qquad \forall \, (r,\phi)\in\mathcal D\,.
\end{equation}}
\end{itemize}
 
\subsection{Rigged Hilbert spaces and $W$-Zernike functions}

The set of $W$-Zernike functions  forms an orthonormal basis for $L^2(\mathcal D)$ so that for any square integrable function $f(r,\phi)\in L^2(\mathcal D)$ we have that
\begin{equation}\label{214}
f(r,\phi)=\sum_{u,v=0}^\infty f_{u,v}\,W_{u,v}(r,\phi)\,,
\end{equation}
with
\begin{equation}\label{215}
f_{u,v}= \int_0^{2\pi} d\phi \int_0^1 dr\,r\, W_{u,v}^*(r,\phi)\,f(r,\phi)\,.
\end{equation}

Let us define two different spaces, which will be the spaces of test functions for respective RHS. The first one is
\begin{equation}\label{216}
\Psi_1:= \left\{ f(r,\phi)\in L^2(\mathcal D) \;\;\Big| \;\; \sum_{u,v=0}^\infty |f_{u,v}|^2\,(u+v+1)^{2p} <\infty\,,\;\; p=0,1,2,\dots \right\}
\end{equation}
The space $\Psi_1$ is endowed with the Fr\`echet topology given by the following family of norms
\begin{equation}\label{217}
||f(r,\phi)||_p^2:= \sum_{u,v=0}^\infty |f_{u,v}|^2\,(u+v+1)^{2p} <\infty\,,\quad p=0,1,2,\dots\,.
\end{equation}
The second space of test functions is defined by the following condition:
\begin{equation}\label{218}
\Psi_2:= \left\{ f(r,\phi)\in L^2(\mathcal D) \;\;\Big| \;\; \sum_{u,v=0}^\infty |f_{u,v}|\,(u+v+1)^{q} <\infty\,,\;\; q=0,1,2,\dots \right\}\,.
\end{equation}
Its topology is given by the following sequence of norms:
\begin{equation}\label{219}
||f(r,\phi)||_{1,q}:= \sum_{u,v=0}^\infty |f_{u,v}|\, (u+v+1)^q\,,\quad q=0,1,2,\dots\,.
\end{equation}

Let us consider a sequence of complex numbers $\{a_n\}$ such that the series $\sum_{n=0}^\infty |a_n|<\infty$. Clearly,
\begin{equation}\label{220}
\sqrt{\sum_{n=0}^\infty |a_n|^2} \le \sum_{n=0}^\infty |a_n|\,,
\end{equation}
which shows that
\begin{equation}\label{221}
\big|\big|f(r,\phi)\big|\big|_p=\sqrt{\sum_{u,v=0}^\infty |f_{u,v}|^2\,(u+v+1)^{2p}} \le \sum_{u,v=0}^\infty \big|f_{u,v}\big|\, (u+v+1)^p= \big|\big|f(r,\phi)\big|\big|_{1,p}\,,
\end{equation}
for $p=0,1,2,\dots$. This shows that $\Psi_2\subset \Psi_1$ and that the canonical injection $i:\Psi_2\longmapsto \Psi_1$ is continuous. This gives a couple of rigged Hilbert spaces where injections in all inclusions are continuous:
\begin{equation}\label{222}
\Psi_2\subset \Psi_1\subset L^2(\mathcal D) \subset \Psi_1^\times \subset \Psi_2^\times\,.
\end{equation} 

An important property for the span of the functions $f(r,\phi)\in\Psi_2$ in terms of the $W$-Zernike functions is given by the following result:\medskip

{\bf Theorem 3.-}{\it 
For any $f(r,\phi)\in\Psi_2$, the series
\begin{equation}\label{223}
f(r,\phi) = \sum_{u,v}^\infty f_{u,v}\,W_{u,v}(r,\phi)\,,
\end{equation}
converges absolutely and uniformly and hence point-wise.
}\smallskip

{\bf Proof.-} 
The proof is based on the bound \eqref{213} valid for the $W$-Zernike functions. Thus, using \eqref{213} and taking into account \eqref{218}, we have that
\begin{equation}\label{224}
\sum_{u,v=0}^\infty \big|f_{u,v}\big|\,\big|W_{u,v}(r,\phi)\big| \le \sum_{u,v=0}^\infty \big|f_{u,v}\big|\, \sqrt{\frac{u+v+1}{\pi}} \le \frac 1{\sqrt\pi} \sum_{u,v=0}^\infty \big|f_{u,v}\big|\, (u+v+1)<\infty\,.
\end{equation}
Then, the Weiersstrass $M$-Theorem guarantees the absolute and uniform convergence of
the series.
\hfill$\blacksquare$

\subsection{Continuity of relevant operators acting on the $W$-Zernike functions}

In the discussion on the continuous basis below, we shall see the relevance of the following operator on 
$L^2(\mathcal D)$:
\begin{equation}\label{225}
P\,f(r,\phi)=r\,e^{i\phi}\,f(r,\phi)\,. 
\end{equation}
In \cite{CGO4}, we prove that
\begin{equation}\label{226}
P\, W_{u,v}(r,\phi) =\alpha_{u}^v\, W_{u+1,v}(r,\phi) + \beta_u^v\,W_{u,v-1}(r,\phi)\,,
\end{equation}
with
\begin{equation}\label{227}
\alpha_u^v = \frac{u+1}{\sqrt{(u+v+1)(u+v+2)}}\,, \qquad \beta_u^v = \frac{v}{\sqrt{(u+v)(u+v+1)}}\,.
\end{equation}
Note that $0\le \alpha_u^v,\beta_u^v \le 1$ and $f_{-1,0}=0$. We want to show that $P\Psi_2\subset \Psi_2$ with continuity. Let us take $f(r,\phi)=\sum_{u,v=0}^\infty f_{u,v}\,W_{u,v}(r,\phi)\in\Psi_2$, so that
\begin{equation*}\begin{array}{lll}\label{228}
\Big|\Big|P \sum_{u,v=0}^\infty f_{u,v}\,W_{u,v}(r,\phi) \Big|\Big|_{1,r} &=&\ds \sum_{u,v=0}^\infty \Big| \alpha_{u-1}^v\,f_{u-1,v}+ \beta_u^{v+1}\,f_{u,v+1} \Big| (u+v+1)^r \\[0.4cm]
& \le&\ds \sum_{u,v=0}^\infty \big|f_{u-1,v}\big| (u+v+1)^r + \sum_{u,v=0}^\infty \big|f_{u,v+1}\big| (u+v+1)^r\,.
\end{array}\end{equation*}
Since $f_{-1,0}=0$, the first term of the second row in \eqref{210} gives
\begin{equation}\begin{array}{lll}\label{229}
\ds \sum_{u,v=0}^\infty |f_{u-1,v}| (u+v+1)^r &=&\ds \sum_{u,v=0}^\infty |f_{u,v}| (u+v+2)^r \le 2^r \sum_{u,v=0}^\infty |f_{u-1,v}| (u+v+1)^r  \\[0.4cm] 
& = &\ds 2^r \Big|\Big| \sum_{u,v=0}^\infty f_{u,v}\,W_{u,v}(r,\phi) \Big|\Big|_{1,r}\,.
\end{array}\end{equation}
The second term in the same row gives,
\begin{equation}\begin{array}{lll}\label{230}
\ds \sum_{u,v=0}^\infty |f_{u,v+1}| (u+v+1)^r &\le&\ds \sum_{u,v=0}^\infty |f_{u,v}|(u+v)^r \le \sum_{u,v=0}^\infty |f_{u,v}|(u+v+1)^r \\[0.4cm] 
&=&\ds \Big|\Big| \sum_{u,v=0}^\infty f_{u,v}\,W_{u,v}(r,\phi) \Big|\Big|_{1,r}\,.
\end{array}\end{equation}
Equations \eqref{229} and \eqref{230} together show that
\begin{equation}\label{231}
\Big|\Big| P \sum_{u,v=0}^\infty f_{u,v}\,W_{u,v}(r,\phi) \Big|\Big|_{1,r} \le (2^r+1) \,\Big|\Big| \sum_{u,v=0}^\infty f_{u,v}\,W_{u,v}(r,\phi) \Big|\Big|_{1,r}\,,
\end{equation}
which shows our claim. 

Other important operators are the generators of the Lie algebra $su(1,1)\oplus su(1,1)$, $U$, $V$, $ A_\pm$, $ B_\pm$,  $A_3=U+1/2$ and $B_3=V+1/2$. Their commutation relations are the following:
\begin{equation}\begin{array}{lll}\label{232}
[U,A_\pm]=\pm A_\pm \,,\quad &[V,B_\pm]=\pm B_\pm\,, \quad &[A_+,A_-]=-2A_3\,,  \\[0.4cm]
 [A_3,A_\pm] =\pm A_\pm\,, \quad &[B_+,B_-]= -2B_3\,, \quad &[B_3,B_\pm]=\pm B_\pm\,.
\end{array}\end{equation}
All the $A$ operators commute with all the $B$ operators. The Casimirs are
\begin{equation}\label{233}
C_A=A_3^2-\frac12 \{A_+,A_-\}\,, \qquad C_B= B_3^2 -\frac 12 \{B_+,B_-\}\,,
\end{equation}
with $\{X,Y\}=XY+YX$. On the $W$-Zernike functions, all these operators act as follows \cite{CGO4}:
\begin{equation}\begin{array}{llllll}\label{234}
U\,W_{u,v}(r,\phi) &=& u\, W_{u,v}(r,\phi)\,,\qquad &V\,W_{u,v}(r,\phi) &=& v\, W_{u,v}(r,\phi)\,,\\[0.4cm]
A_+\, W_{u,v}(r,\phi) &=&(u+1)\,W_{u+1,v}(r,\phi) \,,\qquad &A_-\,W_{u,v}(r,\phi)&=&u\, W_{u-1,v}(r,\phi)\,, \\[0.4cm] 
B_+\,W_{u,v}(r,\phi) &=&(v+1)\,W_{u,v+1}(r,\phi)\,, \qquad &B_-\,W_{u,v}(r,\phi) &=&v\,W_{u,v-1}(r,\phi)\,.
\end{array}\end{equation}
All these operators are densely defined and unbounded on $L^2(\mathcal D)$. Furthermore, 

{\bf Proposition 2}{\it 
The operators $U$, $V$, $ A_\pm$ and $ B_\pm$ are continuous  on $\Psi_2$. In addition, $A_+$ and $A_-$ are formal adjoint of each other and same for $B_+$ and $B_-$ and $U$ and $V$ are essentially self-adjoint on $\Psi_2$. 
}

{\bf Proof}
That $A_+$ and $A_-$ and also $B_+$ and $B_-$ are formal adjoint of each other is obvious from \eqref{217}. The proof of the continuity on $\Psi_2$ of all these operators is the same. Take for instance $A_+$. The formal action of $A_+$ on $f(r,\phi)\in\Psi_2$ is given by
\begin{equation}\label{236}
A_+\, f(r,\phi) = \sum_{u,v=0}^\infty f_{u,v}\,(u+1)\,W_{u+1,v}(r,\phi)\,.
\end{equation}
Then,
\begin{equation}\begin{array}{lll}\label{237}
\big|\big|A_+\, f(r,\phi)\big|\big|_{1,r} &=&\ds \sum_{u=1,v=0}^\infty |f_{u,v}|^2\,(u+1)(u+v+1)^r  \\[0.4cm]
& \le&\ds  \sum_{u,v=0}^\infty |f_{u,v}| (u+v+1)^{r+1} =||f(r,\phi)||_{1,r+1}\,,
\end{array}\end{equation}
which proves that $A_+\Psi_2\subset \Psi_2$ with continuity. The same for all other operators. Finally, $U$ and $V$ are obviously symmetric on $\Psi_2$ and the ranges of $U\pm iI$ and $V\pm iI$ on $\Psi_2$ are $\Psi_2$ itself, so that $U$ and $V$ are essentially self-adjoint with domain $\Psi_2$. 
\subsection{Continuous bases and RHS}

Let $\mathcal H$ be an arbitrary infinite dimensional separable Hilbert space and $U$ a unitary operator $U:\mathcal H= L^2(\mathcal D)$. As in previous cases, we define $\Phi_i:=U^{-1}\Psi_i$, $i=1,2$  \eqref{222}, and transport the topologies on $\Psi_i$ to $\Phi_i$ by $U^{-1}$. We have a couple of rigged Hilbert spaces in correspondence.
So ,we have the following diagram
\[
\begin{array}{lllllclllll}
&\Psi_2&\subset &\Psi_1&\subset &L^2(\mathcal D)&\subset&\Psi_1^\times &\subset&\Psi_2 ^\times
\\[0.2cm]
  & \hskip-0.65cm  {U^{-1}}\downarrow && \hskip-0.65cm {U^{-1}}\downarrow &&\hskip-0.65cm {U^{-1}}\downarrow&&\hskip-0.65cm {U^{-1}}\downarrow&&\hskip-0.65cm {U^{-1}}\downarrow
\\[0.2cm]
&\Phi_2&\subset &\Phi_1 &\subset  & {\mathcal H} &\subset   &\Phi_1^\times &\subset  &\Phi_2^\times
\end{array}\,.
\]
Nevertheless, our rigged Hilbert space of reference will be here $\Phi_2\subset \mathcal H\subset \Phi_2^\times$. Take any vector $U^{-1}f(r,\phi)=|f\rangle\in\Phi_2$, and for (almost with respect to the Lebesgue measure) each $0\le r \le 1$ and $0\le \phi\le 2\pi$ define the mapping $|r,\phi\rangle $ by  $\langle f|r,\phi\rangle:= f^*(r,\phi) =\langle r,\phi|f\rangle^*$. Clearly, $|r,\phi\rangle$ is linear for each $r$ and $\phi$. In addition, this is continuous so that $|r,\phi\rangle\in\Phi_2^\times$. 
To prove the continuity, note that $U^{-1}$ transport the given topology from $\Psi_2$ to $\Phi_2$. Let $|u,v\rangle := U^{-1}\,W_{u,v}(r,\phi)$ for each $u,v=0,1,2,\dots$. Then, if $|f\rangle=\sum_{u,v=0}^\infty f_{u,v}\,|u,v\rangle$, we have that the norms $||\,|f\rangle||_{1,q}$ defining the topology on $\Psi_2$ are identical to \eqref{219}. Thus, taking into account \eqref{213}, we have
\begin{equation}\begin{array}{lll}\label{239}
\big|\langle f|r,\phi\rangle\big| &= &\ds |\langle r,\phi|f\rangle^*| = |f(r,\phi)| \le \sum_{u,v=0}^\infty \big|f_{u,v}\big|\,
\big |W_{u,v}(r,\phi)\big|
 \le \sum_{u,v=0}^\infty |f_{u,v}|\, \sqrt{\frac{u+v+1}{\pi}} \\[0.4cm]
&\le&\ds \frac 1{\sqrt\pi} \sum_{u,v=0}^\infty |f_{u,v}|\,(u+v+1) = \frac 1{\sqrt\pi}\,\big|\big|\,|f\rangle\big|\big|_{1,1}\,.
\end{array}\end{equation}
The scalar product of two vectors $|f\rangle\,,|g\rangle\in\Phi_2$ is given by
\begin{eqnarray}\label{240}
\langle f|g\rangle = \int_0^{2\pi} d\phi\int_0^1 dr \,r \,  f^*(r,\phi)\,g(r,\phi) =  \int_0^{2\pi} d\phi\int_0^1 dr \,r \, \langle f|r,\phi\rangle\langle r,\phi|g\rangle\,,
\end{eqnarray}
so that, we have the identity,
\begin{equation}\label{241}
\mathcal I:= \int_0^{2\pi} d\phi\int_0^1 dr \,r \, |r,\phi\rangle\langle r,\phi|\,,
\end{equation}
which should be interpreted as the canonical injection $\mathcal I:\Phi_2\longmapsto \Phi_2^\times$.  In particular, if we apply \eqref{241} to $|u,v\rangle$, we have that
\begin{equation}\label{242}
|u,v\rangle = \int_0^{2\pi} d\phi\int_0^1 dr \,r \, |r,\phi\rangle\langle r,\phi|u,v\rangle = \int_0^{2\pi} d\phi\int_0^1 dr \,r \, |r,\phi\rangle\,W_{u,v}(r,\phi)\,,
\end{equation}
which may be looked as a relation between the discrete basis $\{|u,v\rangle\}$ in $\mathcal H$ and the continuous basis $\{|r,\phi\rangle\}$. 
Note that, according to our definition, $\langle r,\phi|u,v\rangle =W_{u,v}(r,\phi)$. If we multiply \eqref{242} to the left by $\langle r,\phi|$, we have:
\begin{eqnarray}\label{243}
\langle r',\phi'|u,v\rangle =W_{u,v}(r',\phi') = \int_0^{2\pi} d\phi\int_0^1 dr \,r \, \langle r',\phi'|r,\phi\rangle\,W_{u,v}(r,\phi)\,,
\end{eqnarray}
so that,
\begin{equation}\label{244}
\langle r',\phi'|r,\phi\rangle  = \frac 1r\,\delta(r-r')\,\delta(\phi-\phi')\,.
\end{equation}
Relation \eqref{244} suggest an inversion formula for \eqref{242}. As $\{|u,v\rangle\}$ is an orthonormal basis for $\mathcal H$, we may write the identity on $\mathcal H$ as
\begin{equation}\label{245bis}
 \mathbb I=\sum_{u,v=0}^\infty |u,v\rangle\langle u,v|\,.
\end{equation}
 As $|u,v\rangle\in\Phi_2$, we may write
\begin{equation}\label{245}
|r,\phi\rangle = \sum_{u,v=0}^\infty |u,v\rangle\langle u,v|r,\phi\rangle = \sum_{u,v=0}^\infty |u,v\rangle\,W^*_{u,v}(r,\phi)\,.
\end{equation}
This inversion formula is totally consistent as one may check by formal multiplication to the left by $\langle r',\phi'|$ and the comparison of the given result with \eqref{244} in one side and \eqref{212} on the other. In conclusion, each $|f\rangle\in\Phi_2$ admit two different expansions in terms of the discrete basis $\{|u,v\rangle\}$ and the continuous basis $\{|r,\phi\rangle\}$. They are, respectively,
\begin{equation}\label{246}
|f\rangle =  \sum_{u,v=0}^\infty |u,v\rangle\langle u,v|f\rangle = \sum_{u,v=0}^\infty |u,v\rangle\,f_{u,v}\,,
\end{equation}
and
\begin{equation}\label{247}
|f\rangle =  \int_0^{2\pi} d\phi \int_0^1 dr\,r\, |r,\phi\rangle\langle r,\phi|f\rangle = \int_0^{2\pi} d\phi \int_0^1 dr\,r\, |r,\phi\rangle \,f(r,\phi)\,.
\end{equation}

As a final remark, all operators \eqref{232} have their counterparts as operators on $\mathcal H$ with exactly the same properties. In particular, they are continuous on $\Phi_2$. 

\section{Concluding remarks}\label{conclusions}

Specific RHS are constructed starting from well defined special functions  and a particular UIR of a Lie group, which is the symmetry group of the corresponding special functions.  The Lie generators of these group are continuous operators with the topologies carried by the RHS.

It is a general property that in a RHS the variables and the parameters are one-to-one related.
This implies that, starting from special functions with $n_p$ parameters and $n_v$ continuous variables, it is possible to construct different RHS's. Indeed when $n_p=n_v$ we can construct not only a RHS involving all parameters and variables but also RHS's  involving subsets of equal number of parameters and variables, saving the role of spectators for the remaining ones. If $n_p > n_v$ the possible RHS's  are limited to $n_v$ and the exceeding parameters remain spectators (as it happens with $j$ in Section \ref{su2associatedlaguerre} and $\alpha$ in Section \ref{su11laguerre}) but
it is impossible to construct a RHS based on the $\Gamma(z)$ functions where we have not parameters at all.
An alternative is shown by the Spherical Harmonics where a new variable $\phi$ is added
to the Associated Legendre polynomials, by the extension of Jacobi polynomials to the Jacobi functions defined on the hypersphere ${\mathcal S}^3$ in the subsection \ref{hyperesfere} and by the generalization of Zernike polynomials defined on the interval to Zernike functions defined on the unit circle in Section \ref{zernikeRHS}.

Special functions are transition matrices between discrete and continuous bases (for instance, generalization of the exponential 
$e^{{\bf i}m \phi} = \langle m| \phi\rangle$ in Section \ref{so2preliminary} and spherical harmonics $Y_{l}^{m}(\theta,\phi) = \langle l,m|\theta,\phi\rangle$ in Section \ref{so32sphericalharmonics}).

The UIR of the corresponding  Lie group defines the basis vectors of the discrete basis in the space $\Phi$, while the regular representation of the Lie group defines the basis vectors of the continuous basis in $\Phi^x$ of the RHS $\Phi\subset {\mathcal H}\subset\Phi^\times$.

Special functions determine a basis in the related space of square integrable functions. As they define a basis also of  a unitary irreducible representation of the group, all other bases of the space are simply obtained applying on them an arbitrary element of the group.


\section*{Acknowledgments}
This research is supported in part by the Ministerio de Econom\'ia y Competitividad of Spain  under grant  MTM2014-57129-C2-1-P and the Junta de Castilla y Le\'on (Project BU229P18).






\end{document}